\documentclass[12pt,a4paper]{article}
\pdfoutput=1
\usepackage{fullpage}
\usepackage{jheppub}
\usepackage{parskip}
\usepackage{color}
\usepackage{amsmath}
\usepackage{amsfonts,amssymb}
\usepackage{slashed}            % for slashed characters in math mode
\usepackage{bbm}                % for \mathbbm{1} (unit matrix)
\usepackage{xspace}				% For spacing after commands
\usepackage{tensor}
\usepackage{mathrsfs}
\usepackage{kbordermatrix}
\renewcommand{\kbldelim}{(}
\renewcommand{\kbrdelim}{)}
\usepackage[utf8]{inputenc}

\usepackage{tikz}
\usepackage{tkz-euclide}
\usetkzobj{all}
\usetikzlibrary{decorations.pathmorphing}	% To create gluons and vector bosons
\usetikzlibrary{positioning}
\tikzset{
    v/.style={decorate, decoration={snake, segment length=3mm, amplitude=0.75mm}, draw},
    f/.style={draw=black, postaction={decorate},
        decoration={markings,mark=at position .6 with {\arrow[very thick]{latex}}}},
    fb/.style={draw=black, postaction={decorate},
        decoration={markings,mark=at position .4 with {\arrowreversed[very thick]{latex}}}},
    fnar/.style={draw=black},
    g/.style={decorate, draw=black,
        decoration={coil,amplitude=3pt, segment length=3.5pt}},
    s/.style={dashed,draw=black, postaction={decorate},
        decoration={markings,mark=at position .55 with {\arrow[very thick]{latex}}}},
    sb/.style={dashed,draw=black, postaction={decorate},
        decoration={markings,mark=at position .55 with {\arrowreversed[draw=black,very thick]{latex}}}},
    snar/.style={dashed,draw=black,line width =1.25pt},
}

\newcommand{\cn}[1]{\chi_{#1,L}^{-}} % Negative charginos
\newcommand{\cp}[1]{\chi_{#1,R}^{c,+}} %Conjugate (positive) charginos
\newcommand{\nl}[1]{\chi_{#1,L}^{0}} % Left handed neutralinos
\newcommand{\nr}[1]{\chi_{#1,R}^{c,0}} % Right handed neutralinos

\begin{document}
\title{Probing a slepton Higgs on all frontiers}

\author[a]{Carla Biggio,}
\emailAdd{biggio@ge.infn.it} 
\author[b]{Jeff Asaf Dror,}
\emailAdd{ajd268@cornell.edu} 
\author[b]{Yuval Grossman,}
\emailAdd{yg73@cornell.edu} 
\author[b]{and Wee Hao Ng}
\emailAdd{wn68@cornell.edu} 

\affiliation[a]{Dipartimento di Fisica, Universit\`a di Genova \& INFN, Sezione di Genova, \\via Dodecaneso 33, 16159 Genova, Italy}
\affiliation[b]{Department of Physics, LEPP, Cornell University, Ithaca, NY 14853\\}

\abstract{We study several aspects of supersymmetric models with a $U(1)_R$ symmetry where the Higgs doublet is identified with the superpartner of a lepton. We derive new, stronger bounds on the gaugino masses based on current measurements, and also propose ways to probe the model up to scales of $\mathcal{O}(10\, \textrm{TeV})$ at future $e^+e^-$ colliders. Since the $U(1)_R$ symmetry cannot be exact, we analyze the effects of $R$-symmetry breaking on neutrino masses and proton decay. In particular, we find that getting the neutrino mixing angles to agree with experiments in a minimal model requires a UV cutoff for the theory at around $ 10 \text{ TeV} $.}

 \arxivnumber{}
\maketitle

\section{Introduction}
The LHC collaborations have recently discovered the Higgs boson at around 125 GeV~\cite{Aad:2012tfa,Chatrchyan:2012xdj}, but have yet to find any of the particles which should have appeared below the TeV scale as required to solve the hierarchy problem~\cite{Craig:2013cxa}. 
 This suggests that if supersymmetry (SUSY) is present at the TeV scale, it deviates from its most naive implementations.
There are many suggestions as to how Nature could be supersymmetric but still avoid the bounds applied by the LHC. 
In particular, there has been a lot of interest in substituting the $R$-parity of the Minimal Supersymmetric Standard Model (MSSM) for a continuous $R$-symmetry, ($U(1)_R$)~\cite{Randall:1992cq,Hall:1990hq} (see~\cite{Kribs:2007ac,Benakli:2008pg,Amigo:2008rc,Benakli:2011vb,Benakli:2011kz,Heikinheimo:2011fk,Davies:2011jsm,Kumar:2011np,Fok:2012fb,Kalinowski:2011zzb,Frugiuele:2011mh,Chakraborty:2013gea,Morita:2012kh,Bertuzzo:2012su,Kalinowski:2011zz,Diessner:2014ksa,DeLopeAmigo:2011koa,Kribs:2013eua,Benakli:2012cy,Csaki:2013fla,Fok:2010vk} for recent work in this direction).

One interesting feature of imposing a $U(1)_R$ symmetry 
is that it allows the ordinary down-type Higgs to be 
in a supermultiplet with one of the charged-lepton doublets,
\begin{equation} 
H \equiv \left( H, \ell_L  \right) 
\end{equation} 
and still avoid phenomenological bounds. This intriguing possibility has been discussed in several recent papers: see~\cite{Riva:2012hz, Frugiuele:2011mh,Bertuzzo:2012su} for model building,~\cite{Chakraborty:2015wga} for stop phenomenology, and~\cite{Berger:2015qra} for a suggested explanation of the recent $ eejj , e \nu jj $ excess~\cite{Khachatryan:2014dka,CMS:2014qpa} as well as further discussion on light squark phenomenology. For the purpose of this work we will focus on the possibility that the Higgs doublet is identified with the selectron doublet, though much of our discussion will be more general. This is motivated in section~\ref{sect:model-intro} as it naturally explains the smallness of the electron mass. 

While it is more economical to construct SUSY models where the Higgs is identified with a slepton, usually this causes phenomenological difficulties due to violation of lepton number. In particular, the K\"{a}hler potential generates electroweak-scale Dirac masses between the partner neutrino (defined as the neutral fermionic component of $L_e$) and the gauginos. As a result the partner neutrino generically becomes too heavy. This problem can be avoided by introducing a global symmetry to forbid Majorana neutralino masses, and adding additional adjoint chiral superfields as Dirac partners of the gauginos. This ensures a massless physical neutralino that can be identified with the neutrino. However, due to the smallness of neutrino masses, it is important that the symmetry be preserved under electroweak symmetry breaking. This requires that the global symmetry be an $R$-symmetry such that the neutrino be charged under the $U(1)_R$ but still leave the Higgs uncharged. 

One may wonder why there aren't additional constraints from the many experiments probing lepton flavor number violation. This is because these models generically only have lepton number violation for one flavor (in our case the electron). The stringent limits from lepton number changing processes rely on violation of at least two lepton flavor numbers (most notably $ \mu \rightarrow e \gamma $, which requires muon and electron number violation). 

In this work we explore how Higgs-as-slepton models can be further probed in several different ways. A generic feature of these models is a mixing between the electron doublet and the gauginos, resulting in the physical electron doublet no longer equal to the corresponding gauge eigenstate. This mixing puts bounds on the size of the wino and bino masses. Previous papers have emphasized the corresponding bounds from the high energy frontier through neutral and charged current universality measurements. In this work we explore the limits from low energy measurements of $G_F$. We find these to be more stringent then the high energy constraints for bounds on the bino masses and competitive with bounds on the wino masses. Furthermore, we look at the discovery potential of the future $e^+ e^-$ collider program. Intriguingly, we find that such a machine has the potential to probe this variant of supersymmetry up to ${\cal O} ( 10 \text{ TeV}$).

Another aspect of the model which we will examine is the breaking of $R$-symmetry through Planck-scale effect, naturally generating a small parameter in the theory. This is responsible for generating neutrino masses which would otherwise be zero, but may also lead to effects such as proton decay. 

Experimentally, there has recently been significant development in the neutrino sector. The differences in the squares of the neutrino masses and the three neutrino mixing angles have been measured~\cite{Agashe:2014kda}. Having the Higgs be part of a supermultiplet with the lepton has crucial implications in terms of neutrino phenomenology, the consequences of which we will explore. Planck-scale suppression of $R$-symmetry breaking effects lead to naturally small neutrino masses. Assuming this is the only source of neutrino masses, we find that in order to obtain the large mixing angles measured by neutrino oscillation experiments, the model typically requires a low cutoff scale of at most ${\cal O} (10 \mbox{ TeV})$. In other words, a generic minimal supersymmetric model with the Higgs playing the role of a slepton requires a low ultraviolet (UV) completion scale. 

In addition to contributing to neutrino masses, $R$-symmetry breaking can also lead to proton decay if the gravitino mass is very heavy. Neutrino mass measurements suggests a gravitino mass range between ${\cal O}(10 \text{ eV}) - {\cal O}(10 \text{ keV})$ assuming generic gravity-mediated $R$-breaking. With such masses the model could have rapid proton decay which restricts the possible UV completions of the model.

This paper is structured as follows. We begin by outlining general properties of the Higgs-as-slepton models in Sec.~\ref{sect:model-intro}. We then proceed to study the constraints on gaugino masses from the lepton-gaugino mixing in Sec.~\ref{sect:current-limits}. Phenomenological implications on future $e^+ e^-$ colliders are covered in Sec.~\ref{sect:discovery-potential}. Implications of the lepton mixing angles on these models are discussed in Sec.~\ref{sect:pmns}. We move on to bounds on the gravitino mass from proton decay and neutrino mass measurements in Sec.~\ref{sect:proton-decay}. We conclude in Sec.~\ref{sect:conclusions} with a summary of our main results.

\section{The basics of Higgs-as-slepton models} \label{sect:model-intro}

We consider the most minimal version of the Higgs-as-slepton model from a bottom-up perspective, in which the only additional fields added to the Higgs-less Standard Model (SM) and their supersymmetric partners are the Dirac partners of the gauginos. Table~\ref{tab:model-summary} lists the superfields and their gauge and $U(1)_R$ representations. As mentioned earlier we have chosen the Higgs to be in $L_e$. In places where we generalize our discussion to other choices of lepton flavor, this will be stated in the text. The $R$-charges are chosen so that left-handed (LH) and right-handed (RH) quarks and leptons form $R$-symmetric Dirac pairs, and that the Higgs vacuum expectation value (VEV) does not break $R$-symmetry. 

Note that we keep $B$ and $L$ as free parameters, and thus they are not identified with the usual baryon and lepton numbers. Based on our assignments, the quarks have $R$-charges $B$, the muon and tau $-L$, while the electron always carries $R$-charge $ - 1 $. Moreover, the normalization of $L$ and $B$ is not determined such that different normalization result in different models with different phenomenology. We learn that $B$ and $L$ are parameters that determine the $R$-charge of the quarks and the second- and third-generation lepton superfields.
No significant change in phenomenology arises from different choices of $B$, except for $B =1/3$ or $1$ which lead to rapid proton decay and are hence forbidden (see Sec.~\ref{sect:proton-decay}). Therefore, in our discussion we only consider the generic $B$ case. On the other hand, viable models can be built for several choices of $L$. In particular we will consider the $L=-1$, $L=0$, $L=1$ and the generic $L$ case, that is $L \ne -1,0,1$. Each of these four choices result in distinct lepton phenomenology and hence can be regarded as a separate model.

\begin{table}
\begin{center} 
\begin{tabular}[t]{| c | c | c|}
\hline
 & $ ( SU (3) _C, SU(2) _L ) _Y $  & $U(1)_R$\\
\hline
$H \equiv L_e$ & $(1,2)_{-1/2}$ & $0$\\
$E_e^c$ & $(1,1)_1$ & $2$\\
$L_{\mu,\tau}$ & $(1,2)_{-1/2}$ & $1-L$\\
$E_{\mu,\tau}^c$ & $(1,1)_1$ & $1+L$\\
$Q_{1,2,3}$ & $(3,2)_{1/6}$ & $1+B$\\
$U_{1,2,3}^c$ & $(\bar{3},1)_{-2/3}$ & $1-B$\\
$D_{1,2,3}^c$ & $(\bar{3},1)_{1/3}$ & $1-B$\\
$W^{a\alpha}$ & $(8,1)_0 + (1,3)_0 + (1,1)_0$ & $1$\\
$\Phi^{a}$ & $(8,1)_0 + (1,3)_0 + (1,1)_0$ & $0$\\
\hline
\end{tabular}
\end{center}
\caption{Superfields in the minimal low energy model with the Higgs doublet identified with the selectron doublet. The $U(1)_R$ charges are parameterized with two unknown variables $L$ and $B$, which gives the most general assignment consistent with the requirement of the existence of Yukawas, $R$-charge conservation after electroweak symmetry breaking, and supersymmetry. The $U(1)_R$ in the table refers to the scalar component of the superfield.}
\label{tab:model-summary}
\end{table}

For a generic assignment of $B$ and $L$, the superpotential consistent with the symmetries is
\begin{equation}
\mathcal{W} = \sum_{i,j = 1}^3 y_{d,ij} H Q_i D_j^c + \sum_{i,j \in \{\mu, \tau\}} y_{e,ij} H L_i E_j^c \,.
\end{equation}
For the  $B = 1/3$ or $L = 1$ cases there are extra terms, but we do not discuss them here. In the case $L = 1$, the details of which can be found in~\cite{Bertuzzo:2012su,Chakraborty:2015wga}.

The Higgs-as-slepton model faces a number of difficulties and here we discuss two of them. First is the fact that supersymmetry forbids a mass term for the up-type quarks.  This problem can be solved by introducing non-renormalizable SUSY-breaking K\"{a}hler terms suppressed by a UV cutoff scale, $ \Lambda $,
\begin{equation} 
\int \,d^4\theta \frac{ X ^\dagger }{ M } \frac{ H ^\dagger Q U }{ \Lambda } \,,
\end{equation} 
where $M$ is the $R$-symmetric mediation scale and $X$ is the spurion whose vacuum expectation value $ \left\langle F _X \right\rangle $ corresponds to the SUSY breaking scale. Perturbativity of the couplings requires the cutoff scale to be at most $ 4\pi \text{ TeV} $. Thus the model requires a low-scale UV completion. In principle, one can avoid this by introducing an additional pair of Higgs doublets~\cite{Frugiuele:2011mh,Bertuzzo:2012su}, which then allows top masses to be generated by the tree-level superpotential. However, as we will show in section~\ref{sect:pmns}, reproducing the correct lepton mixing angles also requires a low cutoff if we assume neutrino masses arise from generic $ R $-breaking. This requirement holds even with the additional Higgs doublets. 
The second problem is that the superpotential cannot provide a mass term for the fermion component of the $H=L_e$ doublet (related to the left-handed electron field) since $ H H = 0 $.
Again, this can be resolved by generating a mass in an analogous way~\cite{Grant:1999dr},
\begin{equation} 
\int \,d^4\theta \frac{ X^\dagger X }{ M^2 } \frac{ H D^\alpha H D_\alpha E_3 }{ \Lambda^2 } \,,
\end{equation} 
where $ D_\alpha $ is the superspace derivative. If the electron doublet is the Higgs partner, then this provides a natural explanation for the smallness of the electron mass, hence motivating our original choice.

One of the most important consequences of having the Higgs as a slepton is the mixing between the electroweak gauginos and the Higgs fermionic superpartner. This puts generic constraints on such models. The K\"{a}hler potential generates weak scale Dirac mass terms given by
\begin{equation} 
\int d^4\theta H ^\dagger e ^{ V } H   \supset -\frac{ g v }{ \sqrt{2} } e _L \tilde{W} ^+ - \frac{ g v }{ 2 } \nu_e \tilde{W} ^0 + \frac{ g' v }{ 2} \nu_e \tilde{B} ^0 \,,
\end{equation}
where, $ g $, $ g' $ are the $ SU(2) _L  $ and $ U(1)_Y $ coupling constants and $  v \simeq 246 \text{ GeV}$ is the vacuum expectation value of the Higgs. The Dirac wino and bino masses, $M_{\tilde{W}}$ and $M_{\tilde{B}}$, are of order of the soft $R$-symmetric SUSY-breaking scale $M_{\text{soft}} \equiv \langle F_X \rangle /M$. This implies a mixing of order of the ratio of the electroweak scale to the soft $R$-symmetric scale, which we quantify using the small parameter 
\begin{equation} 
\epsilon \equiv \frac{g v}{2M_{\tilde{W}}} = \frac{m_W}{M_{\tilde{W}}},
\end{equation}
where $ m _W $ is the mass of the $ W $ boson. The above implies that the mass of the gauginos must be high. As discussed in the following, the upper bounds on $ \epsilon $ are $\mathcal{O} 0.1$. The mixing can also depend on the size of the non-renormalizable operators arising at the scale $ \Lambda $. These contributions are model dependent and will be assumed to be negligible. We have also neglected any $ R $-symmetry breaking effects, although we will need to include them when discussing neutrino masses and proton decay later. We also assume that $\vert M_{\tilde W}^2 - M_{\tilde B}^2 \vert \gg m_W^2$. While the mixing between the winos and the binos is modified should we relax this assumption, it turns out to have no significant effects on the phenomenology considered in our work. With the above assumptions, and working to $ {\cal O} ( \epsilon^2 )$ the mass eigenstates are
\begin{align} 
& \cn{1} = \left( 1 -  \epsilon ^2 \right)  e _L ^-  - \sqrt{2}  \epsilon \psi _{\tilde W} ^-  && \cp{1} =  e _R ^{c,+}  \\ 
& \cn{2} = \left( 1 - \frac{1}{2} \epsilon ^2 \right) \psi _{\tilde W} ^- +  \sqrt{2} \epsilon e _L ^-  && \cp{2} = \tilde{W} ^+    \\ 
& \cn{3} = \tilde{W} ^-  &&\cp{3} = \tilde{\psi} ^+ 
\end{align} 
\begin{align} 
& \nl{1} = \left( 1 - \frac{1}{2} \epsilon ^2  \left( 1 + \alpha ^2 t _w ^2 \right) \right)  \nu _e - \epsilon \psi  _{\tilde W} + \epsilon \alpha t _w \psi _{\tilde B} && \\ 
& \nl{2} = \left( 1 - \frac{1}{2} \epsilon ^2 \right) \psi _{\tilde W} + \epsilon \nu _e + \epsilon ^2 \frac{ \alpha t _w }{ 1 - \alpha }  \psi _{\tilde B} && \nr{2} = \tilde{W} ^0 + \epsilon ^2 \frac{ \alpha ^2 t _w }{ 1 - \alpha ^2 }   \tilde{B} \\ 
& \nl{3} = \left( 1 - \frac{2}{1} \epsilon ^2 \alpha ^2 t _w ^2 \right) \psi _{\tilde B}  - \epsilon \alpha t _w \nu _e - \epsilon ^2 \frac{ \alpha ^3 t _w }{ 1 - \alpha ^2 }  \psi _{\tilde W} &&\nr{3} = \tilde{B}  - \epsilon ^2 \frac{ \alpha ^2 t _w }{ 1 - \alpha ^2 } \tilde{W} ^0 
\end{align} 
where $t_w$ denotes the tangent of the Weak mixing angle, and $ \alpha \equiv M _{\tilde W} / M _{\tilde B} $. (For details on the mixing matrices and diagonalization, see appendix~\ref{app:FeynmanRules}.)

\section{Limits on gaugino-electron doublet mixing} \label{sect:current-limits}
Previous works have shown that the strongest constraints on the model arise from the mixing between the gaugino and the electron doublet~\cite{Riva:2012hz,Frugiuele:2011mh}. The bounds from neutral current universality have been emphasized (with a mention of the weak charged-current universality bounds in Ref.~\cite{Riva:2012hz}). Charged-current interactions also provide a different set of constraints through non-standard neutrino interactions (NSI)~\cite{Biggio:2009nt,Davidson:2003ha,Barranco:2007ej,Ohlsson:2012kf,Bolanos:2008km,Mitsuka:2011ty}. In this section we compute the neutral-current bounds in our general framework and compare the results with additional bounds from NSI. Note that at tree-level neutral current effects can only constrain the wino masses since this arises from mixing of the electrons in the $ Z  e e $ interaction, while charged current measurements are affected by both electron and neutrino mixing in the $ W e \nu $, yielding bounds on both the wino and bino masses.

We start by computing the electron neutral current. Definitions of the mixing matrices $U_{C,L}$, $U_{C,R}$ and $U_{N,L}$ used here are provided in Appendix~\ref{app:FeynmanRules}. The interaction is given by
\begin{align} 
 \Delta {\cal L} &= \frac{ g }{ c _w } \bigg[ \left( c _w ^2 - \left| ( U _{ C,R} ) _{ 11} \right| ^2 \right) ( \cp{1}) ^\dagger \bar{\sigma} ^\mu Z_\mu \cp{1} \notag \\ 
& \hspace{2cm}- \left( c _w ^2 - \frac{1}{2} \left| ( U _{ C,L} ) _{ 11} \right| ^2 \right) ( \cn{1} ) ^\dagger \bar{\sigma} ^\mu Z _\mu  \cn{1} \bigg].   
\end{align} 
Keeping only terms to $ {\cal O} ( \epsilon ^2  ) $, this gives
\begin{equation} 
\Delta {\cal L} = \frac{ g }{ c _w } \left[ - s _w ^2 ( \cp{1}) ^\dagger \bar{\sigma} ^\mu \cp{1} - \left( \frac{1}{2} - s _w ^2 \right) ( \cn{1} ) ^\dagger \bar{\sigma} ^\mu \cn{1}  \right] Z _\mu - \frac{ g }{ c _w } \epsilon ^2 ( \cn{1} ) ^\dagger \bar{\sigma} ^\mu Z _\mu  \cn{1},
\end{equation} 
from which we obtain the axial current coupling of the $Z$ to fermions
\begin{equation} 
g _A = g_A^{SM} \left[ 1 + 2 \epsilon ^2 \right] , \qquad g_A^{SM}=\frac{ g }{ 2 c _w },
\end{equation} 
where $g_A^{SM}$ is the SM value of the axial coupling.
(Bounds on the vector current are much weaker and hence irrelevant for this discussion.)
Experimentally the bounds on the axial current are~\cite{Agashe:2014kda},
\begin{equation} 
\left| \frac{ \delta g _A ^e }{ g _A ^{ e}} \right| \approx 1.2 \times 10 ^{ - 3} \quad (90\%\text{ CL} ) \,.
\end{equation} 

This stringent bound applies only to the wino mass. Bounds on the bino mass arise from modifications of the charged current. The left-handed electron charged current are described by
\begin{align} 
\Delta {\cal L} & = g \left( ( U _{ N,L} ) ^\ast _{ 21} ( U _{ C,L} ) _{ 21} + \frac{1}{\sqrt{2}} ( U _{ N,L} ) ^\ast _{ 11} ( U _{ C,L} ) _{ 11} \right) W _\mu ( \cp{1} )  ^\dagger   \bar{\sigma} ^\mu \nl{1} \\ 
& = \frac{ g }{ \sqrt{2} } \left(1 + \frac{ \epsilon ^2 }{ 2} \left( 1 - \alpha ^2 t _w ^2 \right) \right) W _\mu ( \cp{1} )  ^\dagger   \bar{\sigma} ^\mu \nl{1}\,.
\end{align} 
Ref.~\cite{Riva:2012hz} computed the charged current universality constraints from $ \tau $ decays. This corresponds to the limit~\cite{Loinaz:2004qc},
\begin{equation} 
\frac{ \left| \delta g \right|    }{ g ^{ SM} } \lesssim 2.6 \times 10 ^{ - 3} \quad (90\% \text{ CL}) \,.
\end{equation} 
There are more stringent constraints arising from NSI interactions. The most stringent constraint, in models where the Cabibbo-Kobayashi-Maskawa (CKM) matrix is assumed to be unitary, arise from taking the ratio of $ G _F $ measured in two different ways. The first is through beta- and Kaon- decays and the second (and more precise) through muon decay. If the CKM is unitary then these should be equal to each other and the ratio gives the bound~\cite{Biggio:2009nt},
\begin{equation} 
\frac{ \left| \delta g \right|    }{ g ^{ SM} } \lesssim 4.0 \times 10 ^{ - 4} \quad (90\% \text{ CL}) \,.
\end{equation} 

\begin{figure} 
  \begin{center} 
\includegraphics[width=8cm]{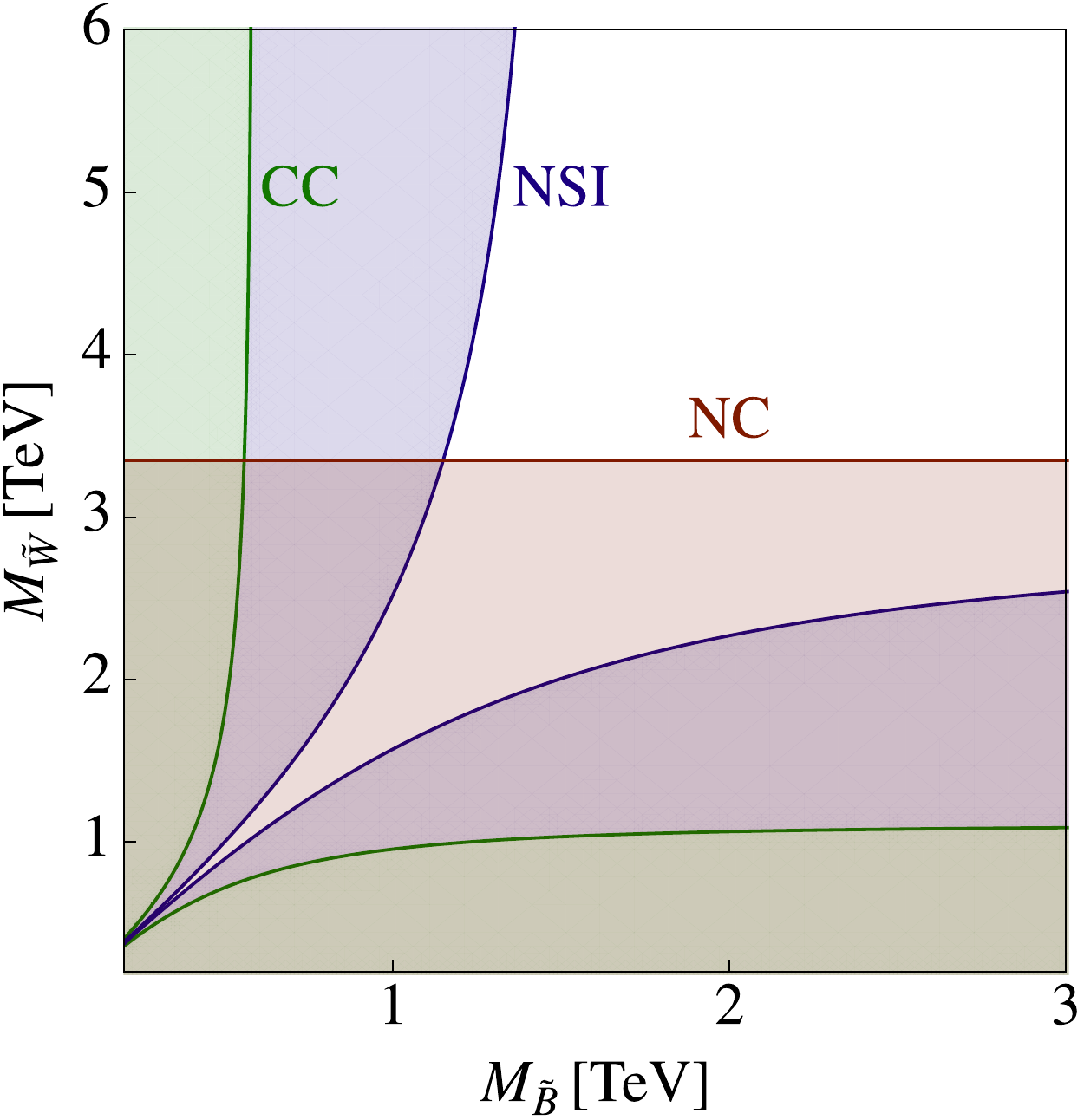} 
\caption{Current limits on the bino and wino masses. The regions in blue are excluded by NSI constraints and depend on both the bino and wino mass, while the region in red is ruled out by neutral current constraints. The limits from charged current universality are shown in green.}
\label{fig:limit}
\end{center}
\end{figure}

This limit, as well as the one from the neutral current, are presented in figure~\ref{fig:limit}. We see that while neutral current interactions place a stronger constraint on the wino mass than NSI, it does not constrain the bino mass. 
Meanwhile, the NSI bounds on the bino mass are generally weaker than on the wino mass due to a $t_w$ suppression in the bino mixing with the neutrino. Combining the NSI and neutral current bounds, we can put a constraint on the bino mass of $M_{\tilde{B}} \gtrsim 1.2 \text{ TeV} $. This is more stringent than the existing universality constraint of about $ 500 \text{ GeV} $~\cite{Riva:2012hz}.

\section{Discovery potential at an $ e ^+ e ^- $ collider} \label{sect:discovery-potential}
\begin{figure} 
\begin{center} 
\begin{tikzpicture} [line width= 1.2pt]
\draw[fb] (0,0) node[below,left]{$ e $} -- (0.5,.75);
\draw[f] (0,1.5) node[above,left]{$ e $}  -- (0.5,0.75);
\draw[v](0.5,.75) -- (1.5,.75) node[midway,above] {$ \gamma $} % node[below,midway] {$ q \rightarrow $}
;
\draw[v] (1.5,.75) -- (2,1.5) node[right]{$ W ^-   $}; %\node[] at (4,1.5) {$ k _1  $}; \node[] at (4.4,1.9) {$ \nearrow $}; 
\draw[v] (1.5,.75) -- (2,0) node[right] {$ W ^+  $};
\begin{scope}[shift={(3.5,0)}]
\draw[fb] (0,0) node[below,left]{$ e $} -- (0.5,.75);
\draw[f] (0,1.5) node[above,left]{$ e $}  -- (0.5,.75);
\draw[v](0.5,.75) -- (1.5,.75) node[midway,above] {$ Z $};% node[below,midway] {$ q \rightarrow $};
\draw[v] (1.5,.75) -- (2.,1.5) node[right]{$ W ^-  $}; %\node[] at (4,1.5) {$ k _1  $}; \node[] at (4.4,1.9) {$ \nearrow $}; 
\draw[v] (1.5,.75) -- (2,0) node[right] {$ W ^+  $};
\end{scope}
\begin{scope}[shift={(7,0)}]
\draw[fb] (0,0) node[below,left]{$ e $} -- (1,0);
\draw[f] (0,1.5) node[above,left]{$ e $}  -- (1,1.5);
\draw[f](1,1.5) -- (1,0) node[midway,right] {$ \nu  $};% node[left,midway] {$ \tilde{q} $}; %\node[] at (1.25,0.5) {$\downarrow$};
% node[below,midway] {$ q \rightarrow $};
\draw[v] (1,1.5) -- (2,1.5) node[right]{$ W ^-  $}; %\node[] at (4,1.5) {$ k _1  $}; \node[] at (4.4,1.9) {$ \nearrow $}; 
\draw[v] (1,0) -- (2,0) node[right] {$ W ^+  $};
\end{scope}

\begin{scope}[shift={(1.5,-2.5)}]
\draw[fb] (0,0) node[below,left]{$ e $} -- (1,0);
\draw[f] (0,1.5) node[above,left]{$ e $}  -- (1,1.5);
\draw[fb](1,0) -- (1,1.5) node[midway,right] {$ e$};
\draw[v] (1,0) -- (2,0) node[above, right]{$ Z $};
\draw[v] (1,1.5) -- (2,1.5) node[below, right] {$ Z$};

\begin{scope}[shift={(3.5,0)}]
\draw[fb] (0,0) node[left]{$ e $} -- (1,0);
\draw[f] (0,1.5) node[left]{$ e $}  -- (1,1.5);
\draw[fb](1,0) -- (1,1.5) node[midway,left] {$ e$};
\draw[v] (1,0) -- (2,1.5) node[right]{$ Z $};
\draw[v] (1,1.5) -- (2,0) node[ right] {$ Z$};
\end{scope}
\end{scope}
\begin{scope}[shift={(0,-5)}]
  \draw[fb] (0,0) node[below,left]{$ e $} -- (1,0);
\draw[f] (0,1.5) node[above,left]{$ e $}  -- (1,1.5);
\draw[fb](1,0) -- (1,1.5) node[midway,right] {$ \tilde{\chi}  _2 ^- $};
\draw[snar] (1,0) -- (2,0) node[above, right]{$ h $};
\draw[v] (1,1.5) -- (2,1.5) node[below, right] {$ Z$};

\begin{scope}[shift={(3.5,0)}]
\draw[fb] (0,0) node[below,left]{$ e $} -- (1,0);
\draw[f] (0,1.5) node[above,left]{$ e $}  -- (1,1.5);
\draw[fb](1,0) -- (1,1.5) node[left,midway] {$ \tilde{ \chi} _2 ^-  $};
\draw[v] (1,0) -- (2,1.5) node[right]{$ Z $};
\draw[snar] (1,1.5) -- (2,0) node[below, right] {$ h$};
\end{scope}

\begin{scope}[shift={(7,0)}]
\draw[fb] (0,0) node[left] {$ e $} -- (0.5,0.75);
\draw[f] (0,1.5) node[left] {$ e $} -- (0.5,0.75);
\draw[v] (0.5,0.75) -- (1.5,0.75) node[midway,above] {$Z$}; 
\draw[v] (1.5,0.75) -- (2,1.5) node[right] {$Z$}; 
\draw[snar] (1.5,0.75) -- (2,0) node[right] {$h$}; 
\end{scope}
\end{scope}
\end{tikzpicture}
\end{center}
\caption{Feynman diagrams for the $ 2 \rightarrow 2 $ processes that we consider in this work. The top row shows $ e ^+ e ^-  \rightarrow W ^+ W ^-  $, the middle row represents $ e ^+ e ^- \rightarrow Z Z $, and the bottom process is $ e ^+ e ^- \rightarrow Z h $. We use $ \tilde{\chi} _2 ^- $ to denote the Dirac spinor $ \left( \cn{2} , (\cp{2}) ^\dagger \right) $.}
\label{fig:feyn}
\end{figure}
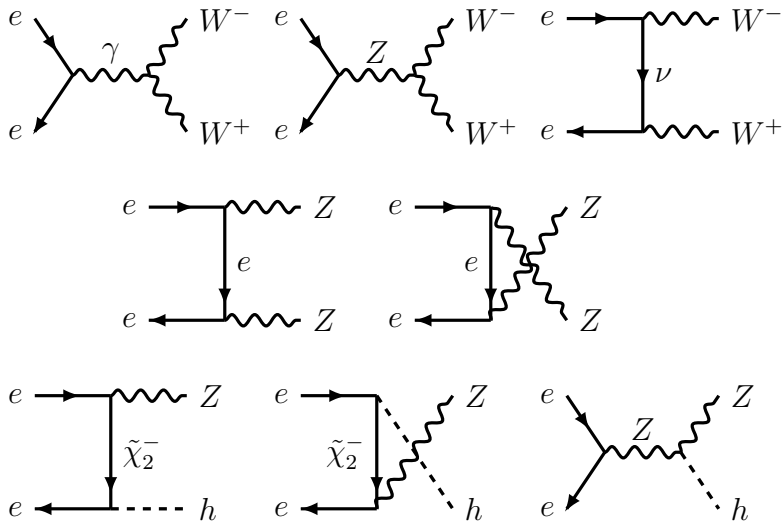

The Higgs-as-slepton model generates deviations of the SM couplings in the electron interactions through modifications of pure SM couplings and from additional interactions with the gauginos. This leads us to expect significant discovery potential at an $ e ^+ e ^- $ collider. In this section we consider different $ 2 \rightarrow 2 $ processes that will deviate from their SM predictions. In the following we keep terms to $ {\cal O} ( \epsilon ^2 ) $ and we ignore all non-renormalizable corrections arising at the scale $ \Lambda $. In particular we consider, $ e ^+ e ^- \rightarrow W ^+ W ^- , ZZ, h Z $. The relevant Feynman diagrams are displayed in figure~\ref{fig:feyn}. Naively one would expect to also have $ e ^+ e ^- \rightarrow h h $ arising from chargino exchange, however these turn out not to arise at tree level up to $ {\cal O} ( \epsilon ^4 ) $ due to angular momentum conservation suppressing $ s $-wave production. 
We use the Feynman rules detailed in Appendix~\ref{app:FeynmanRules} to compute the cross-sections.

To study projections at a future collider we use the condition that the significance, that we take to be $S / \sqrt{ B}$, where $ S $ is the signal and $B$ is the background, is larger than $1.645$ (corresponding to a 90\% confidence interval),
\begin{equation} 
\frac{ {\cal L} \times   \delta \sigma   }{ \sqrt{ {\cal L} \times \sigma ^{ SM}}} > 1.645 \,,
\end{equation} 
where 
$ {\cal L} $ is the luminosity of the collider and $ \delta \sigma \equiv \sigma ^{ BSM} -  \sigma ^{ SM} $. We expect this to be a reasonable estimate due to the controlled environment offered by a lepton collider, leading to negligible backgrounds. 

One subtlety is the cross-section diverges for small $ t $, or equivalently small $ \left| \eta \right| $, due to a Rutherford singularity. In order to remove sensitivity to this divergence we cut off the phase space integration at $ \left| \eta \right| = 2 $. To avoid this complication in our expressions, we quote the differential $ d (\delta  \sigma)/dt  $ for each process. 

\subsection{$ e ^+ e ^- \rightarrow W ^+ W ^-$}
We begin by computing the effects to $ e ^+ e ^-  \rightarrow W ^+ W ^- $ scattering. The Feynman diagrams which contribute up to $ {\cal O} ( \epsilon ^2  ) $ are shown in figure~\ref{fig:feyn}. Note that there are no diagrams with virtual charginos or neutralinos since adding these requires paying the price of additional $ \epsilon $'s in the vertices. The only modifications to the SM cross-section are from deviations in the $ Z e e $ couplings. The effects considered here are a close analog to deviations considered in $ t W \rightarrow tW $ scattering at the LHC from anomalous $ Ztt $ couplings~\cite{Dror:2015nkp}. The cross-sections are straightforward to compute but the expressions are complicated without making some approximations. For simplicity we only quote the result to lowest order in $ m _V ^2 / s $ ($ V = h,Z, $ or $ W $), though in producing the figures we use the full expressions. 
The result for the signal is
\begin{equation} 
\frac{ d (\delta \sigma)  }{ d t }= \frac{1}{4} \frac{\beta }{32 \pi  s} \left\{ \frac{ 2 e ^4 }{ s _w ^4 } \frac{ \left(  \frac{1}{2} - s _w ^2 - \alpha ^2 \right) }{ \alpha ^2 }\frac{ ( - t )  }{ M _{\tilde W} ^2 }\frac{   s + t  }{ s }  + {\cal O} \left( \frac{ m _V ^2 }{ s   } ,\frac{ s ^2  }{ M _{\tilde W} ^4  }\right) \right\}  \,,
\end{equation} 
where $ \beta \equiv \sqrt{ 1 - 4 m _W ^2  /s }  $ is the velocity of either $ W $ boson, $ s _w $ is the sine of the weak mixing angle, $ s \equiv ( p _{e ^{ - }} + p _{ e ^+ } ) ^2 $,  $ t \equiv ( p _{ W ^- } + p _{ W ^+ } ) ^2 $ and $ \alpha = m _Z / m _W $.

\subsection{$ e ^+ e ^- \rightarrow Z Z $}
Next we consider $ e ^+ e ^- \rightarrow ZZ $ scattering, depicted in figure~\ref{fig:feyn}. As for $ W ^+ W ^- $, the chargino-exchange diagrams only arise at higher orders in $ \epsilon $. Also in this process the deviation from the SM is in the $ Z e e $ coupling, but, unlike in the $ W ^+ W  ^- $ case the total cross-section does not grow with energy but is roughly constant. The difference of the energy scaling between $ ZZ $ and $ W ^+ W ^- $ production can be traced back to the algebra of $ SU(2) $ or equivalently the fact that there doesn't exist a triple gauge coupling $ ZZZ $ in the model. The signal is, 
\begin{equation} 
\frac{d( \delta  \sigma )}{ dt } = \frac{1}{2} \frac{1}{4} \frac{ \beta }{ 32 \pi s } \left\{ \frac{ 2 e ^4 }{ s _w ^4 c _w ^4 } \left( 1 - 2 s _w ^2 \right) ^2 \frac{ m _W ^2 }{ M _{\tilde W} ^2 } \frac{ s ^2 + 2 s t + 2 t ^2  }{ t ( s + t ) }  + {\cal O} \left( \frac{ m _V ^4 }{ s ^2  } ,\frac{ s ^2  }{ M _{\tilde W} ^4  }\right)  \right\} \,,
\end{equation} 
where here $ \beta \equiv \sqrt{ 1 - 4 m _Z ^2 /s } $ gives the speed of one of the $Z$ bosons. 

Note that for $ e ^+ e ^- \rightarrow ZZ $ the deviation of the coupling is factorizable as the two diagrams (see figure~\ref{fig:feyn}) have the same dependence on the anomalous coupling. Thus the new physics contribution is just a rescaling of the SM cross-section. 

\subsection{$ e ^+ e ^- \rightarrow h Z $}
Another interesting channel at a lepton collider is $ h Z $ production. The Feynman diagrams are shown in figure~\ref{fig:feyn} with the beyond the SM (BSM) effects entering from chargino exchange as well as modifications to the $ Zee $ coupling. Since the $ \tilde{\chi} _2 h e $ vertex does not have an $ \epsilon $ suppression, these diagrams are still of $ {\cal O} ( \epsilon ^2 ) $. The signal is,
\begin{equation} 
\frac{ d(\delta  \sigma) }{ dt }  = \frac{1}{4} \frac{ \beta }{ 32 \pi s } \left\{ \frac{ e ^4 }{ s _w ^4 c _w ^2  } \left( \frac{1}{2}   -   s _w ^2 \right)  \frac{ ( - t ) }{ M _{\tilde W} ^2 } \frac{ s + t }{ s}  + {\cal O} \left( \frac{ m _V ^2 }{ s ^2  } ,\frac{ s ^2  }{ M _{\tilde W} ^4  }\right)  \right\} \,,
\end{equation} 
where 
\begin{equation}
\beta \equiv \sqrt{ 1 - m _Z ^2 / E _Z ^2 }, \qquad E_Z \equiv \frac{ \sqrt{s} }{ 2} \left( 1 + \frac{ m _Z ^2 }{ s} - \frac{ m _h ^2 }{ s ^2 } \right)
\end{equation}
such that $\beta$
denotes the speed of the $Z$ boson. The signal is roughly the same as that of $ W ^+ W ^- $ production, however the SM cross-section of $ h Z $ is significantly smaller due to the relatively small $ h ZZ $ vertex. This makes deviations easier to identify, increasing its sensitivity to new physics.

\begin{figure}
\begin{center} 
\includegraphics[width=10cm]{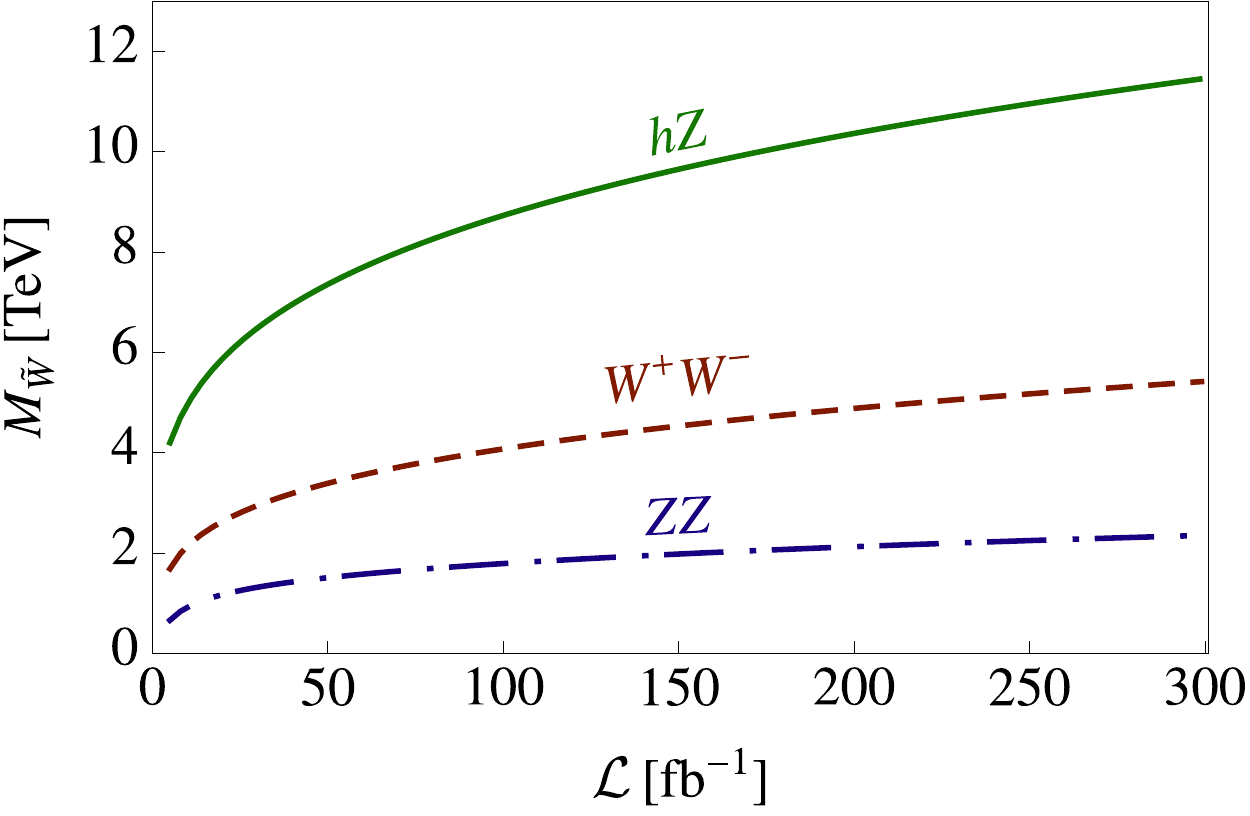} 
\caption{The potential reach from $ e ^+ e ^- \rightarrow V V $ at a future lepton collider as a function of luminosity. The $ h Z $ deviations are by far the largest as they scale quickly with energy and have suppressed SM contributions compared to $ W ^+ W ^- $.}
\label{fig:reach}
\end{center} 
\end{figure}

Figure~\ref{fig:reach} compares the reach of the different channels as a function of luminosity for a $ 1 \text{ TeV} $ linear collider. 
The reach at such a collider is striking. A $ 300 \text{fb} ^{-1} $ collider can probe wino masses up to $ M _{\tilde W} \sim 5.4 \text{ TeV} $, $M _{\tilde W} \sim  2.3 \text{ TeV}  $, and $M _{\tilde W} \sim 11.5 \text{ TeV} $ for $ W ^+ W ^- , ZZ, $ and $  h Z $ respectively. The scale probed by $ h Z $ is impressive, exploring physics well beyond the TeV scale. Furthermore, correlated excesses in all these channels would be a smoking-gun for the model. These projections highlight the promising opportunities offered by an $e ^+ e ^- $ collider in testing Higgs-as-slepton models.

Lastly, we note that three body production channels can likely be used to probe the model further. In particular, modifications to $ h h Z $ production (important for measuring the Higgs-trilinear coupling) are also affected at $ {\cal O} ( \epsilon ^2 ) $. We leave the study of these channels for future work.

\section{$U_{\text{PMNS}}$ and the need for a TeV-scale cutoff} \label{sect:pmns}

We next discuss the neutrinos sector in Higgs-as-slepton models. For 
a generic choices of $L$, that is, $L \ne 0,1,-1$, the $U(1)_R$ symmetry forbids neutrino masses. Thus, all neutrino masses are $U(1)_R$-breaking, which can naturally explain the hierarchy between neutrinos and the rest of SM fermions masses. (Exceptions occur in the case $L=0,-1$, which we will address later.) One extra ingredient in the model is that since it singles out one neutrino flavor to be the Higgs superpartner, this can lead to suppression of the mixing between the Higgs-partner neutrino with the other two neutrino flavors, with obvious implications for the Pontecorvo-Maki-Nakagawa-Sakata (PMNS) matrix, $U_{\text{PMNS}}$. A large suppression of one or more of the mixing angles would be inconsistent with measured values of the $|\theta_{12}| \approx 0.6$, $|\theta_{23}| \approx 0.7$ and $|\theta_{13}| \approx 0.15$ \cite{Agashe:2014kda}.

In this section, we show that for generic gravity-mediated $U(1)_R$-breaking, consistency with the measured mixing angles requires that the cutoff-scale $\Lambda$ be less than ${\cal O}( 10 \mbox{ TeV})$, so that non-renormalizable contributions to the neutrino mass matrix be of comparable sizes to that from mixing-induced contributions involving gaugino soft masses. This turns out to be the case regardless of the choice of $L$. It is interesting to note that the upper-bound on the cutoff scale is similar to the one required for generating a large enough top quark mass, despite the two phenomena being  unrelated. While not the focus of this work, we also briefly discuss neutrino mixing in the Higgs-as-slepton model with two additional Higgs doublets (in principle this can replace the UV cutoff needed to produce the top mass). We find that such models also generically require a low energy cutoff, except for  particular choices of $ L $. 

\subsection{$L \ne -1,0,1$}

We establish our analysis framework using the $L \ne -1,0,1$ case as an example. We first derive the $3 \times 3$ neutrino mass matrix from the full neutralino mass matrix, which we then use to obtain the mixing angles required to diagonalize the neutrino mass matrix. We assume generic gravity mediation and we estimate the sizes of the matrix elements using a spurion analysis, assuming ${\cal O}( 1)$ coefficients and including non-renormalizable contributions involving the cutoff $\Lambda$. Measured values of the mixing angles then translate to bounds on $\Lambda$.

To provide a useful picture of the mass scales involved, we refer to Sec.~\ref{sect:proton-decay}, where we find that the gravitino mass should be $m_{3/2} \sim {\cal O}( 10 \text{ eV} - 100 \text{ eV} )$ in order to provide the correct neutrino masses. This is much smaller than the $U(1)_R$-symmetric soft mass scale which, as we discussed above, are of the order of few TeVs.

\subsubsection{Neutrino mass matrix}

In Sec.~\ref{sect:current-limits} and \ref{sect:discovery-potential}, where we studied electroweak precision and collider phenomenology, the main effects came from the mixing between the Higgs-partner neutrinos and the gauginos. Therefore, it was convenient to ignore $U(1)_R$-breaking masses and work with Dirac mass matrices, even for the neutralinos. However, since we are now interested in the mixing between neutrino flavors, the $U(1)_R$-breaking masses play an important role and so it is more useful to work with a Majorana mass matrix instead.

We begin with the tree-level $7 \times 7$ neutralino Majorana mass matrix in the interaction basis $\{ \nu_e,\allowbreak \nu_\mu,\allowbreak \nu_\tau,\allowbreak \tilde{B},\allowbreak \tilde{W}^0,\allowbreak \psi_{\tilde{B}},\allowbreak \psi_{\tilde{W}}^0 \}$. We first diagonalize the matrix only with respect to the $U(1)_R$-symmetric terms, from which we find that three of the eigenvectors $\{\nu'_e,\allowbreak \nu_\mu,\allowbreak \nu_\tau\}$ do not have $U(1)_R$-symmetric masses, where $\nu'_e$ is given to order ${\cal O}( \epsilon)$ by
\begin{equation}
\nu'_e \simeq \nu_e + \epsilon t_w \alpha \psi_{\tilde{B}} - \epsilon \psi_{\tilde{W}^0}.
\end{equation}
These three eigenvectors can still have $U(1)_R$-breaking masses. The associated $3 \times 3$ block of the transformed $7 \times 7$ neutralino Majorana mass matrix is (the origin of the terms is derived below)
\begin{equation}
M_\nu \equiv 
\bordermatrix{ & \nu'_e & \nu_\mu & \nu_\tau \cr
\nu'_e         & c_{\psi_{\tilde{W}}} + t_w^2\alpha^2 c_{\psi_{\tilde{B}}} +\epsilon' c_{ee} &\epsilon'  c_{e\mu} & \epsilon' c_{e\tau} \cr
\nu_\mu        & \epsilon' c_{\mu e} & \epsilon' c_{\mu\mu} &\epsilon' c_{\mu\tau} \cr 
\nu_\tau       & \epsilon' c_{\tau e} & \epsilon' c_{\tau\mu} & \epsilon' c_{\tau\tau},
 } \epsilon^2 m_{3/2}
\end{equation}
where
\begin{equation}
\epsilon' \equiv \frac{2M_{\tilde{W}}^2}{g^2\Lambda^2}.
\end{equation}
$\epsilon'$ can be roughly interpreted as the ratio of the soft mass scales to the cutoff scale of the model. Therefore, a small $\epsilon'$ implies a high cutoff scale, while an $\mathcal{O}(1)$ $\epsilon'$ implies a low cutoff scale only slightly above the sparticle masses.

The overall factor of $\epsilon^2 m_{3/2}$ can be understood from the fact that the neutrino masses break both $U(1)_R$ and electroweak symmetry. We now explain the origin of the various mass terms. The first two terms in $(M_\nu)_{ee}$ arise from the fact that $\nu'_e$ contains $\psi_{\tilde{B}}$ and $\psi_{\tilde{W}}^0$, which in turn are involved in the soft $U(1)_R$-breaking neutralino mass terms
\begin{equation}
\int d^4 \theta \frac{X^\dagger}{M_{\text{Pl}}} \left( \frac{c_{\psi_{\tilde{B}}}}{2} \Phi_{\tilde{B}} \Phi_{\tilde{B}} + \frac{c_{\psi_{\tilde{W}}}}{2} \Phi_{\tilde{W}} \Phi_{\tilde{W}} \right)
\supset
m_{3/2} \left( \frac{c_{\psi _{ \tilde{B}}}}{2} \psi_{\tilde{B}} \psi_{\tilde{B}} + \frac{c_{\psi _{ \tilde{W}}}}{2} \psi_{\tilde{W}}^0 \psi_{\tilde{W}}^0 \right)\,,
\end{equation}
where $c_{\psi_{\tilde{B}}}$ and $c_{\psi_{\tilde{W}}}$ are arbitrary ${\cal O}( 1)$ coefficients since we have assumed generic gravity mediation. As for the other matrix elements, they can be generated by non-renormalizable operators of the form
\begin{equation}
\int d^4\theta \frac{X^\dagger}{M_{\text{Pl}}\Lambda^2}\frac{1}{2} 
c_{ij} \left( L_e^\dagger e^V L_i \right) \left(L_e^\dagger e^V L_j \right) 
\supset \frac{M_{\tilde{W}}^2}{\Lambda^2} \frac{c_{ij}}{g^2} \epsilon^2 m_{3/2} \nu_i \nu_j \,,
\end{equation}
where $i,j \in \{ e, \mu, \tau\}$, and we have again assumed $c_{ij}$ to be ${\cal O}( 1)$. Note that we have replaced $v^2$ by $\frac{4M_{\tilde{W}}^2}{g^2} \epsilon^2$ to make the $\epsilon$-dependence manifest.

In principle, one should also take into account loop contributions to $M_\nu$. Generically, we expect the contribution to $(M_{\nu})_{ee}$ to be of order $(\epsilon^2 m_{3/2})/(16\pi^2)$, which is a loop factor smaller than the first two tree-level terms and can hence be systematically ignored. For the other matrix elements, the loop contributions cannot be achieved with a single soft $U(1)_R$-breaking insertion (the soft terms cannot supply the required number of units of $U(1)_R$-breaking for these elements), and so require an insertion of a nonrenormalizable operator, in which case they are also a loop factor smaller than the corresponding tree-level terms. 
Since we will show that agreement with the measured $U_{\text{PMNS}}$ requires a low TeV-scale cutoff $\Lambda$, these loop contributions are definitely much smaller than the corresponding tree-level non-renormalizable contributions and so it is consistent to ignore the former without affecting the validity of our final results.

Finally, we argue that $M_\nu$ should in fact be regarded as the $3 \times 3$ neutrino mass matrix. The neutrino mass matrix is obtained by block-diagonalizing the transformed $7 \times 7$ neutralino mass matrix, this time with respect to the $U(1)_R$-breaking masses. However, since the four other transformed states have masses $M_{\tilde{W}}$ or $M_{\tilde{B}}$, the remaining ``transformation angles'' required for block-diagonalization are at most of ${\cal O}( \frac{\epsilon^2 m_{3/2}}{M_{\tilde{W}}})$ or ${\cal O}( \frac{\epsilon^2 m_{3/2}}{M_{\tilde{B}}})$. This implies that the basis $\{\nu'_e,\allowbreak \nu_\mu,\allowbreak \nu_\tau\}$ is very close to the actual basis required for block-diagonalization, and also that the resulting ``corrections'' to $M_\nu$ are at most  ${\cal O}( \frac{\epsilon^4 m_{3/2}}{M_{\tilde{W}}} m_{3/2})$ or ${\cal O}( \frac{\epsilon^4 m_{3/2}}{M_{\tilde{B}}} m_{3/2})$ and hence negligible.

\subsubsection{Reproducing $ U _{ PMNS} $}

To obtain the mixing angles in $U_{\text{PMNS}}$, we need to find the transformations that diagonalize the charged-lepton and neutrino mass matrices. We first consider the charged-lepton sector. Unlike the neutrinos, the charged-lepton masses are dominated by $U(1)_R$-symmetric contributions. Therefore, the $3 \times 3$ charged-lepton Dirac mass matrix is block-diagonal between the electron and the other lepton flavors to a very good approximation since mass terms of the form $e'_L \mu_R^c$, $e'_L \tau_R^c$, $\mu_L e_R^{\prime c}$ and $\tau_L e_R^{\prime c}$ are $U(1)_R$-breaking and hence much smaller. Therefore, we are completely justified in choosing the lepton flavor basis to coincide with the charged-lepton mass basis, since the required transformation does not involve the Higgs-partner generation. This means that the PMNS mixing angles are entirely determined by the neutrino sector.

We now consider the neutrinos. We first assume that we can have a high cutoff scale so that $\epsilon' \ll 1$\, in which case the neutrino mass matrix takes the form
\begin{equation}
M_\nu \sim  \bordermatrix{ & \nu'_e & \nu_\mu & \nu_\tau \cr
\nu'_e & {\cal O} (1) & {\cal O}( \epsilon')  & {\cal O}( \epsilon') \cr
\nu_\mu & {\cal O}( \epsilon')  & {\cal O}( \epsilon')  & {\cal O}( \epsilon') \cr
\nu_\tau & {\cal O}( \epsilon')  & {\cal O}( \epsilon')  & {\cal O}( \epsilon') } \epsilon^2 m_{3/2} \,.
\end{equation}
We find that the neutrino mass eigenstate $\nu_1$ (associated most closely with $\nu'_e$) is much heavier than $\nu_2$ and $\nu_3$, and that both mixing angles $\theta_{12}$ and $\theta_{13}$ are of order $\epsilon'$ and hence small. These observations are inconsistent with experimental measurements, implying that we cannot have $\epsilon' \ll 1$. Rather, a $\mathcal{O}(1)$ $\epsilon'$ is preferred. In the best-case scenario, allowing for fluctuations in ${\cal O}( 1)$ coefficients, we place a lower bound of $\epsilon' \gtrsim {\cal O}( 0.1)$, which in turn implies that
\begin{equation}
\Lambda \lesssim {\cal O} \left( \frac{\sqrt{20}}{g} M_{\tilde{W}} \right) \,.
\end{equation}
For $ M _{\tilde W} \sim  \text{TeV} $ the required  cutoff scale is ${\cal O}( 10 \mbox{ TeV})$. This ensures that the non-renormalizable contributions to $M_\nu$ are comparable to the mixing-induced gaugino soft-term contributions to $(M_\nu)_{ee}$ which is required to have large neutrino mixing angles and a mass hierarchy consistent with measurements. Note that it is possible to evade the mass hierarchy issue associated with $\epsilon' \ll 1$ by choosing a different lepton generation for the Higgs (e.g. the choice $\tau$ is consistent with normal hierarchy), but the problems associated with the mixing angles remain.

Finally, we recall that in order to generate the top mass in the Higgs-as-slepton model we  require $\Lambda \lesssim {\cal O}( 10 \, \text{TeV} )$. It is interesting to note that both the top mass and neutrino mixing, that are 
 unrelated physical phenomena, both point towards an ${\cal O}( 10 \mbox{ TeV})$ upper bound for the cutoff scale.

\subsection{$L=1$}

Now we consider the case with $L = 1$ where there are two main differences with respect to the general case discussed above. The first is % not so important at is
the fact that in the neutrino sector, the loop contributions to all the $M_\nu$ matrix elements can now be generated by a single soft $U(1)_R$-breaking insertion (whereas this is only true for $(M_\nu)_{ee}$ when $L \ne 1$). Nevertheless, being at least one loop factor smaller than the soft-mass contribution to $(M_\nu)_{ee}$, they are still too small to replace the need for a low cutoff scale $\Lambda$. 

The second effect is more important; in the charged-lepton sector, the mass terms $e'_L \mu_R^c$, $e'_L \tau_R^c$, $\mu_L e_R^{\prime c}$ and $\tau_L e_R^{\prime c}$ are no longer $U(1)_R$-breaking, so the charged-lepton Dirac mass matrix isn't diagonal anymore. If we choose the flavor basis to be the charged-lepton mass basis, it is no longer guaranteed that the Higgs be associated with a single flavor, i.e. all the sneutrinos can in principle get VEVs. On the other hand, such a scenario is inconsistent with bounds on lepton-flavor violating processes such as $\mu \rightarrow e\gamma$ \cite{Agashe:2014kda}. For example, if all the sneutrinos get VEVs, the $W$ and $Z$ gauge coupling vertices will then mix the gauginos with all three charged-lepton mass eigenstates such that a $W/Z$-gaugino loop can induce $\mu \rightarrow e\gamma$. Therefore, any successful implementation of the $L=1$ scenario requires that the sneutrino VEVs be suppressed for two of the generations, which, returning to our original flavor basis, suggests that the Dirac mass matrix should again be approximately block-diagonal. (Note that this also implies that the $L=1$ model is less favorable than the generic $L$ model due to the need for the sneutrino VEV suppression in the other two generations.)

Therefore, we conclude that these differences do not affect our conclusion of the need for a TeV-scale cutoff. We note that the same conclusion was made in \cite{Bertuzzo:2012su} in the context of a Two Higgs Doublet Model (2HDM) extension of the Higgs-as-slepton model. As a result, the authors introduced a right-handed Dirac neutrino as a low-scale UV completion, which is analogous to our idea of a cutoff scale $\Lambda$.

The above discussion is only valid for generic gravity mediated $U(1)_R$-breaking. As discussed in \cite{Bertuzzo:2012su}, anomaly mediation does not generate soft mass terms of the form $\psi_{\tilde{W}^0}\psi_{\tilde{W}^0}$ and $\psi_{\tilde{B}}\psi_{\tilde{B}}$, so in fact the neutrino mass matrix can be entirely dominated by loop contributions without any constraints on $\Lambda$.

\subsection{$L=0$}

For $L=0$, before imposing any additional symmetry, the non-renormalizable contributions to $\nu_\mu \nu_\mu$, $\nu_\mu \nu_\tau$ and $\nu_\tau \nu_\tau$ are no longer $U(1)_R$-breaking. As a result, two of the neutrinos become too heavy. Therefore, for such a choice to work, one needs to impose an additional global $U(1)$ lepton number symmetry on $L_\mu$ and $L_\tau$ \cite{Frugiuele:2011mh}, assumed to be broken at some flavor scale $M_f$. At this scale we get an $ R $-conserving but lepton symmetry-violating operator, %The non-renormalizable contributions now read
\begin{equation}
\int d^4\theta \frac{X^\dagger}{M_f \Lambda^2}\frac{1}{2} 
c_{ij} \left( L_e^\dagger e^V L_i \right) \left(L_e^\dagger e^V L_j \right) 
\supset \alpha' \frac{M_{\tilde{W}}^2}{\Lambda^2} \frac{c_{ij}}{g^2} \epsilon^2 m_{3/2} \nu_i \nu_j \quad (i,j \in \{\mu, \tau\}) \,,
\end{equation}
where $\alpha' \equiv M_{\text{Pl}}/M_f \ge 1$. Note that we have assumed that the $M_f$-scale mediators can also mediate SUSY-breaking, due to the involvement of the spurion $X$. Otherwise, we should either replace one of the $\Lambda$ by $M$, or replace $M_f$ by $M_{\text{Pl}}$, whichever gives the lower overall suppression. As a result, $M_\nu$ now takes the form
\begin{equation}
M_\nu \sim  \bordermatrix{ & \nu'_e & \nu_\mu & \nu_\tau \cr
\nu'_e & {\cal O}( 1) & {\cal O}( \epsilon')  & {\cal O}( \epsilon') \cr
\nu_\mu & {\cal O}( \epsilon')  & {\cal O}( \alpha'\epsilon')  & {\cal O}( \alpha'\epsilon') \cr
\nu_\tau & {\cal O}( \epsilon')  & {\cal O}( \alpha'\epsilon')  & {\cal O}( \alpha'\epsilon') } \epsilon^2 m_{3/2}.
\end{equation}

There are two scenarios that result in the neutrino mixings angles, $\theta_{12}$ and $\theta_{13}$, that are very small, which we would like to avoid. The first is if $\epsilon' \ll 1$, and the second if $\alpha'\epsilon' \gg 1$. To avoid both scenarios, we require that $ \epsilon ' \gtrsim 0.1 $ and $ \alpha ' \epsilon ' \lesssim 10 $ (or equivalently $\alpha' \lesssim 100$). The first constraint again corresponds to a low TeV-scale cutoff as was found in the previous cases. The second constraint corresponds to $ M _f \gtrsim M _{ \text{Pl}} / 100 $ or, in other words, that we need the flavor scale cutoff to be close to the Planck scale so that the $U(1)_R$-symmetric neutrino masses do not become too large. Therefore, the lepton number symmetry should be broken very close to the Planck scale.
Yet, we note that this conclusion assumes that $M_f$-scale mediators can also mediate SUSY-breaking, and is not valid otherwise.

\subsection{$L=-1$}

Next, we consider the $L=-1$ case. While less obvious than the $L=0$ case, we also have the problem of two of the neutrinos becoming too heavy. This can seen from the fact that $\nu_e$, $\psi_{\tilde{W}^0}$ and $\psi_{\tilde{B}}$ have $U(1)_R$-charges $-1$, while $\nu_\mu$, $\nu_\tau$, $\tilde{W}^0$ and $\tilde{B}$ have $U(1)_R$-charges $+1$, so there can be three massive Dirac pairs at the $U(1)_R$-symmetric level, leaving only one massless neutralino. More specifically, one can come up with $U(1)_R$-symmetric mass terms such as
\begin{equation}
\begin{aligned}
&\int d^4 \theta \frac{X^\dagger}{M} \left( c_{\tilde{B}i} \frac{ \Phi_{\tilde{B}} L_e^\dagger e^V L_i }{\Lambda} + c_{\tilde{W}i} \frac{\Phi_{\tilde{W}}^a L_e^\dagger e^V \tau^a L_i}{\Lambda}  + 
c_{ei} \frac{\left( L_e^\dagger e^V L_e \right) \left(L_e^\dagger e^V L_i \right)}{\Lambda^2} \right)\\
&\supset \frac{M_{\text{Pl}}}{M} m_{3/2} \left( \frac{\sqrt{2}M_{\tilde{W}}}{g\Lambda} c_{\tilde{B}i} \epsilon \psi_{\tilde{B}} \nu_i +
\frac{M_{\tilde{W}}}{\sqrt{2} g\Lambda} c_{\tilde{W}i} \epsilon \psi_{\tilde{W}^0} \nu_i + \frac{M_{\tilde{W}}^2}{g^2\Lambda^2} c_{ei} \epsilon^2 \nu_e \nu_i \right),
\end{aligned}
\end{equation}
for $i \in \{\mu,\tau\}$, leading to large neutrino masses. Note that $\frac{M_{\text{Pl}}}{M} m_{3/2}$ gives the soft $U(1)_R$-symmetric scale.

As in the $ L = 0 $ case, one way to resolve this issue is to introduce an additional $U(1)$ lepton symmetry on $L_\mu$ and $L_\tau$, both of which are broken at the flavor scale $M_f$. As a result, all instances of $M$ in the above equation should be replaced by $M_f$. Assuming $M_f$ to be large and hence the above terms to be much smaller than the original $U(1)_R$-symmetric masses, we can then follow the previous procedure to obtain the neutrino mass matrix. In other words, we first diagonalize the full $7\times 7$ Majorana mass matrix with respect to the original $U(1)_R$-symmetric terms, following which we block-diagonalize with respect to the remaining lepton symmetry-breaking and/or $U(1)_R$-breaking terms. We find that $M_\nu$ now takes the form
\begin{equation}
M_\nu \sim  \bordermatrix{ & \nu'_e & \nu_\mu & \nu_\tau \cr
\nu'_e & {\cal O}( 1) & {\cal O}( \alpha'\epsilon')  & {\cal O}( \alpha'\epsilon') \cr
\nu_\mu & {\cal O}( \alpha'\epsilon')  & {\cal O}( \epsilon')  & {\cal O}( \epsilon') \cr
\nu_\tau & {\cal O}( \alpha'\epsilon')  & {\cal O}( '\epsilon')  & {\cal O}( \epsilon') } \epsilon^2 m_{3/2}.
\end{equation}

Again, there are two scenarios that lead to small neutrino mixing(s) which we want to avoid. The first is if $\epsilon' \ll 1$, leading to one or two small angles depending on the size of $\alpha'\epsilon'$. The second is if $\alpha'\epsilon' \gg 1$, leading to one small angle. Therefore, just as in the $L=0$ case, we again see that we require both a low  cutoff-scale $\Lambda$, and a lepton number-breaking scale $M_f$ close to the Planck scale. Note that the constraints here are slightly weaker since the suppression may now occur only for one mixing angle, which can be identified with the smallest measured angle $\theta_{13}$.

\subsection{2HDM Higgs-as-slepton model} \label{sect:2hdm-pmns}
Finally, we discuss the 2HDM Higgs-as-slepton model (see appendix~\ref{app:2hdm} for a summary of the differences), where we will only consider the $L \ne -1,0,1$ case for brevity. The 2HDM model may be one possible UV completion of the Higgs-as-sneutrino model \cite{Riva:2012hz}, completing the model to a much higher scale since the top quark can now gain mass from the up-type Higgs (although the electron mass still has to come from non-renormalizable operators). We now show that the requirement of lepton mixing angles forces also the 2HDM model to have a much lower UV completion scale than one might expect. %\yg{what is ``expected'' here?}

The analysis follows the same procedure as before, although it is now complicated by the fact that there are two additional neutralinos, one associated with the up-type Higgs $\tilde{h}_u^0$, and another with the electroweak doublet required for anomaly cancellation $\tilde{r}_d^0$ (these correspond to the superfields $ H _u $ and $ R _d $). Also, there are now additional soft $U(1)_R$-breaking terms that can contribute to the neutrino mass matrix via mixing. For instance, we can now have
\begin{equation}
\int d^4 \theta \frac{X^\dagger}{M_{\text{Pl}}} c_i L_i H_u
\supset
c_i m_{3/2}  \nu_i \tilde{h}_u^0
\end{equation}
where $i \in \{e,\mu,\tau\}$. This enters the neutrino mass matrix since $\nu'_e$ now also contains a $\tilde{h}_u^0$ component. Finally, being a 2HDM model, there is also a $\tan\beta \equiv v_u / v_d$ dependence (where $v_u(v_d)$ is the vacuum expectation value of $h_u(h_d)$).

We find that the neutrino mass matrix takes the form
\begin{equation}
M _\nu  \sim 
\bordermatrix{ & \nu'_e & \nu_\mu & \nu_\tau \cr
\nu'_e         & {\cal O}(c_\beta^2) + {\cal O}(c_\beta s_\beta) + {\cal O}(\epsilon') & {\cal O}(c_\beta s_\beta) + {\cal O}(\epsilon') & {\cal O}(c_\beta s_\beta) + {\cal O}(\epsilon') \cr
\nu_\mu        & {\cal O}(c_\beta s_\beta) + {\cal O}(\epsilon') & {\cal O}(\epsilon') & {\cal O}(\epsilon') \cr 
\nu_\tau       & {\cal O}(c_\beta s_\beta) + {\cal O}(\epsilon') & {\cal O}(\epsilon') & {\cal O}(\epsilon')
 } \epsilon^2 m_{3/2}
\end{equation}
where $c_\beta \equiv \cos\beta$ and $s_\beta \equiv \sin\beta$.
If we assume that $c_\beta s_\beta \sim {\cal O}(1)$ or $c_\beta^2 \sim {\cal O}(1)$, then we again find one or two mixing angles with size ${\cal O}(\epsilon')$. Therefore, we see that even in the 2HDM model, we still need a low cutoff scale in order to reproduce the PMNS matrix. In general the constraint is slightly weaker than before due to the $\beta$ dependence. This is a non-trivial result since the 2HDM version can otherwise have a much higher cutoff scale given that the top quark mass can be generated by $H_u$ rather than through nonrenormalizable operators. On the other hand, if $t_\beta \gg 1$, we expect both $c_\beta s_\beta$ and $c_\beta^2$ to be small, in which case the constraints on the cutoff scale can be less stringent depending on the size of $t_\beta$. In particular, for large $ t _\beta $ the required cutoff scale is,
\begin{equation}
\Lambda \lesssim \sqrt{\frac{20}{g^2} t_\beta} M_{\tilde{W}}\,,
\end{equation}
raising the cutoff by a factor of $\sqrt{t_\beta}$.

We note that the above conclusion is invalid for the case $L = 0$, since in this specific case the ${\cal O}(\epsilon')$ terms in the lower right $2\times 2$ block are then replaced by ${\cal O}(\alpha'\epsilon')$. A small $\epsilon'$ can be compensated by a large $\alpha'$ to give large mixing angles. In other words, a larger cutoff-scale $\Lambda$ can be compensated for by a smaller flavor scale $M_f$.

\section{Neutrino masses, proton decay and the gravitino mass} \label{sect:proton-decay}
%\subsection{Overview}

The $U(1)_R$ symmetry in Higgs-as-slepton models serves two important roles: to forbid neutrino masses (as long as the gauginos have separate Dirac mass partners $\psi_{\tilde{G}}$, $\psi_{\tilde{W}}$ and $\psi_{\tilde{B}}$) as well as to forbid superpotential and soft terms that might have otherwise led to rapid proton decay. However, since neutrino masses are small but nonzero, we require explicit breaking of the $U(1)_R$ symmetry, possibly through gravity mediation to account for this smallness. In particular, this implies a relation between the neutrino masses and the gravitino mass $m_{3/2} \approx \langle F_X \rangle/M_{\text{Pl}}$, the details of which depends on whether the breaking is through generic ``Planck-scale'' gravity mediation or through anomaly mediation. The $U(1)_R$-breaking may also introduce proton decay channels, which lead to upper bounds on the gravitino mass $m_{3/2}$. It is hence of interest to discuss the bounds on $m_{3/2}$ from the neutrino mass spectrum and from proton decay. In this section we restrict our attention to the case of generic gravity mediation, since the proton decay channels we consider below do not arise in anomaly mediation despite the $U(1)_R$-breaking.

\subsection{Bounds from neutrino masses}

We have already discussed neutrino masses in Sec.~\ref{sect:pmns} and so we will only briefly review the relevant points. If $L \ne -1,0$, then all neutrino masses involve $U(1)_R$-breaking and hence scale with the gravitino mass $m_{3/2}$. In particular, for generic gravity mediation, we have shown that the Majorana mass for the Higgs-partner neutrino is given by $\sim \epsilon^2 m_{3/2}$. This arises mainly from the mixing of the neutrino with $\psi_{\tilde{B}}$ and $\psi_{\tilde{W}}^0$ and is generally larger than loop-induced masses. We use this to set the mass scale of the heaviest neutrino, since all other terms in the neutrino mass matrix are expected to be of the same order so as to explain the large mixing angles in $U_{\text{PMNS}}$. Even for the cases $L = 0$ and $L=-1$, while some of the neutrino mass terms are $U(1)_R$-symmetric, we require them to be suppressed by some flavor scale $M_f$ close to the Planck scale so that these mass terms are comparable to the mixing-induced term above.

\begin{figure} 
\begin{center} 
\includegraphics[width=9cm]{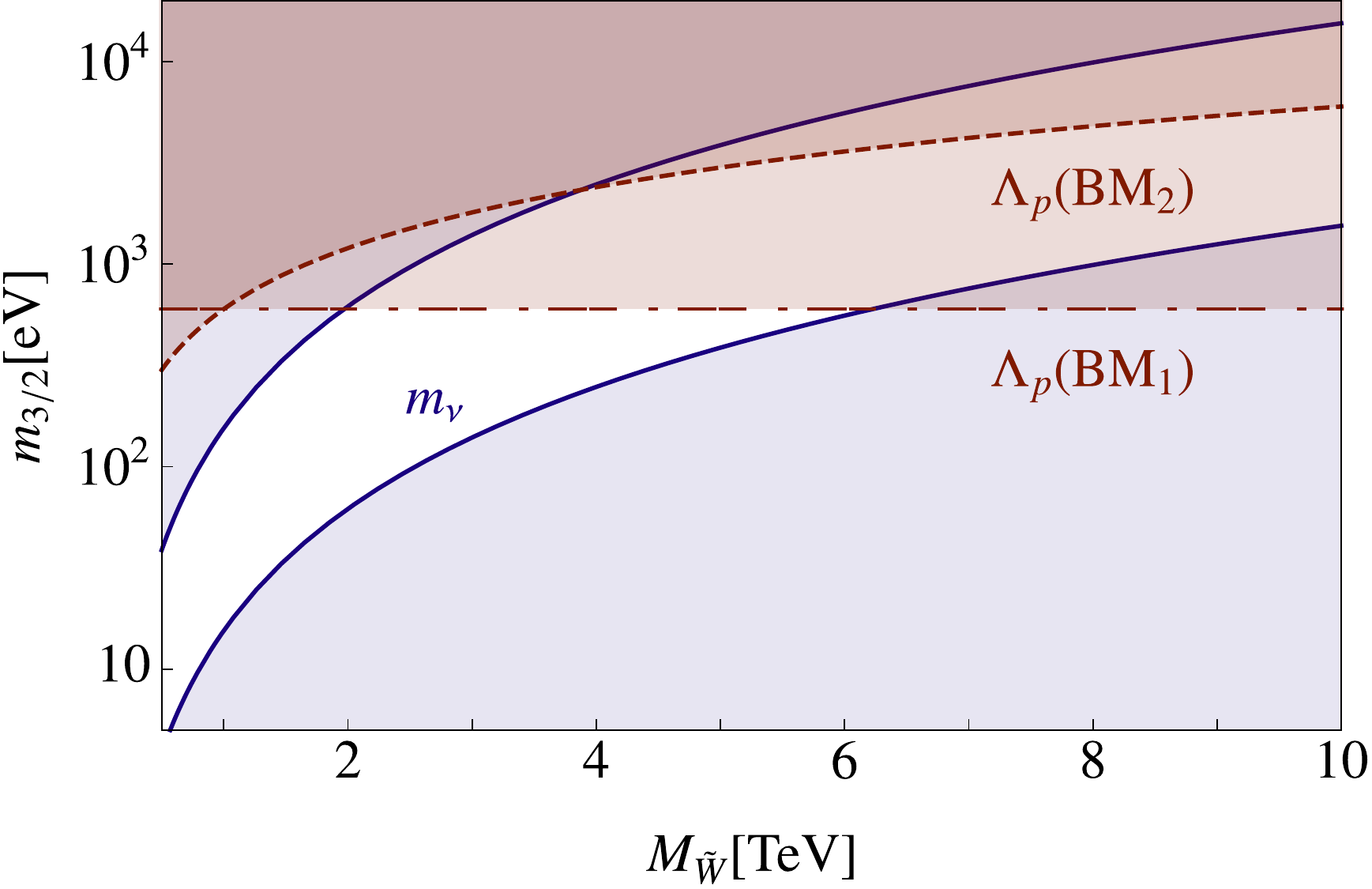} 
\end{center}
\caption{The excluded gravitino mass range. The limits in blue correspond to constraints from the neutrino mass scale while the limits in red are from proton lifetime measurements. The constraints from the proton lifetime are dependent on the $ m _{ \tilde{s} _R } $,$m _{ \tilde{g} } $, and we include two benchmark scenarios. $ \text{BM}_1 $ is for $ m _{ \tilde{s} _R }  =  M _{\tilde g} = 1\text{ TeV} $ while $ \text{BM}_2 $ is for $ m _{ \tilde{s} _R }=1 \text{ TeV} $, $ M _{\tilde g} = M _{\tilde W} $.}
\label{fig:gravitino}
\end{figure}

Mass hierarchy measurements from neutrino oscillation experiments require the heaviest neutrino mass to be at least around $0.1 \, \text{eV}$, while cosmology and spectroscopy experiments place an upper bound of around $1 \, \text{eV}$ \cite{Agashe:2014kda}. Together, this implies the following bounds on the gravitino mass:
\begin{equation}
\left( \frac{0.1}{\epsilon}\right)^2 10 \, \text{eV} \lesssim m_{3/2} \lesssim \left( \frac{0.1}{\epsilon}\right)^2 100 \, \text{eV}.
\end{equation}
Note that the bounds are dependent on the wino mass through $\epsilon$. The allowed values of the gravitino mass are shown in Fig.~\ref{fig:gravitino} as a function of the wino mass, with the excluded region shown in blue.

\subsection{Upper bounds from proton decay}
\label{sec:protondecay}
After generic gravity-mediated $U(1)_R$-breaking, various operators appear that can give rise to proton decay. For example, we now have $a_{ijk} U_{Ri}^c D_{Rj}^c D_{Rk}^c$ in the superpotential, which comes from
\begin{equation}
\mathcal{L} \supset \int d^4 \theta A_{ijk}\frac{X^\dagger}{M_{\text{Pl}}\Lambda} U_{Ri}^c D_{Rj}^c D_{Rk}^c,
\end{equation}
so $a_{ijk} = (m_{3/2}/\Lambda) A_{ijk}$, where $A_{ijk}$ are ${\cal O}( 1)$ coefficients. In conjunction with $y_{d,ij} L_1 Q_{Li} D_{Rj}^c \equiv y_{d,ij} H_d Q_{Li} D_{Rj}^c$ already present in the $U(1)_R$-symmetric superpotential, this gives rise to tree-level proton decay, familiar from the $ R $-parity violating MSSM. Remember that we have already excluded the $B=1$ scenario, in which $a_{ijk} U_{Ri}^c D_{Rj}^c D_{Rk}^c$ is $U(1)_R$-symmetric and hence $a_{ijk}$ is entirely unsuppressed, leading to rapid proton decay.

\begin{figure}[t]
\scalebox{0.95}{
\begin{tikzpicture} 
\draw[f] (0,0) -- (1.5,0) node[pos=0,left] {$u_1$}; 
\draw[f] (0,-3) -- (1.5,-3) node[pos=0,left] {$d_1$};
\draw[g] (1.5,0) -- (1.5,-1.5) node[midway,left=2.5pt] {$\tilde{g}$};
\draw[g] (1.5,-1.5) -- (1.5,-3) node[midway,left=2.5pt] {$\tilde{g}$};
\draw[fb] (1.5,-1.5) -- (1.5,0);
\draw[fb] (1.5,-1.5) -- (1.5,-3);
\draw[fnar] (1.3,-1.3) -- (1.7,-1.7);
\draw[fnar] (1.3,-1.7) -- (1.7,-1.3);
\draw[s] (1.5,0) -- (3,-1.5) node[midway,above=2pt,right] {$\tilde{u}_{R1}^c$};
\draw[s] (1.5,-3) -- (3,-1.5) node[midway,below=2pt,right] {$\tilde{d}_{R1}^c$};
\draw[s] (4.5,-1.5) -- (3,-1.5) node[midway,above=2pt] {$\tilde{d}_{Rk}^c$} node[pos=1,left] {$b_{11k}$};
\draw[fb] (4.5,-1.5) -- (6,0) node[pos=1,right] {$\overline{\nu_e}$} node[pos=0,right=2pt] {$-y_{d,lk}$};
\draw[fb] (4.5,-1.5) -- (6,-3) node[pos=1,right] {$\overline{d_l}$};
\draw[f] (0,-4) -- (6,-4) node[pos=0,left] {$u_1$};
\end{tikzpicture}}
\quad
\scalebox{0.95}{
\begin{tikzpicture} 
\draw[f] (0,0) -- (1.5,0) node[pos=0,left] {$u_1$}; 
\draw[f] (0,-3) -- (1.5,-3) node[pos=0,left] {$d_1$};
\draw[g] (1.5,0) -- (1.5,-1.5) node[midway,left=2.5pt] {$\tilde{g}$};
\draw[g] (1.5,-1.5) -- (1.5,-3) node[midway,left=2.5pt] {$\tilde{g}$};
\draw[fb] (1.5,-1.5) -- (1.5,0);
\draw[fb] (1.5,-1.5) -- (1.5,-3);
\draw[fnar] (1.3,-1.3) -- (1.7,-1.7);
\draw[fnar] (1.3,-1.7) -- (1.7,-1.3);
\draw[s] (1.5,0) -- (3,-1.5) node[midway,above=2pt,right] {$\tilde{u}_{R1}^c$};
\draw[s] (1.5,-3) -- (3,-1.5) node[midway,below=2pt,right] {$\tilde{d}_{R1}^c$};
\draw[s] (4.5,-1.5) -- (3,-1.5) node[midway,above=2pt] {$\tilde{d}_{Rk}^c$} node[pos=1,left] {$b_{11k}$};
\draw[fb] (4.5,-1.5) -- (6,0) node[pos=1,right] {$e^+$} node[pos=0,right=2pt] {$-y_{d,lk} (V_{\text{CKM}})_{lm}^\dagger$};
\draw[fb] (4.5,-1.5) -- (6,-3) node[pos=1,right] {$\overline{u_m}$};
\draw[f] (0,-4) -- (6,-4) node[pos=0,left] {$u_1$};
\end{tikzpicture}}
\caption{One-loop proton decay channels arising from soft trilinear scalar terms $\tilde{u}_{Ri}^c \tilde{d}_{Rj}^c \tilde{d}_{Rk}^c$ and the Majorana gluino mass. All indices here label mass eigenstates. The cross indicates a Majorana gluino mass insertion. There is a similar set of diagrams involving the Majorana mass of the gluino Dirac partner.} \label{fig:proton-decay}
\end{figure}
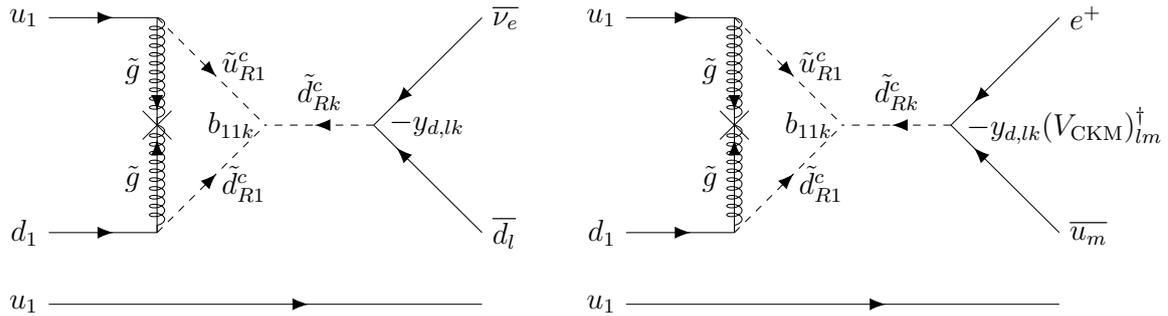

Another possibility is the one-loop proton decay channels shown in Fig.~\ref{fig:proton-decay}, which requires soft trilinear terms $b_{ijk} \tilde{u}_{Ri}^c \tilde{d}_{Rj}^c \tilde{d}_{Rk}^c$, as well as the soft Majorana mass $m_{\tilde{g}}$ and $m_{\psi_{\tilde{g}}}$ for the gluinos and their Dirac partners. The latter are always $U(1)_R$-breaking, so we expect that $m_{\tilde{g}} = c_{\tilde{g}} m_{3/2}$ and $m_{\psi_{\tilde{g}}} = c_{\psi_{\tilde{g}}} m_{3/2}$, where $c_{\tilde{g}}$ and $c_{\psi_{\tilde{g}}}$ are ${\cal O}(1)$ coefficients. For $B\ne 1/3$, the trilinear terms are also $U(1)_R$-breaking, so we expect that $b_{ijk} = B_{ijk} m_{3/2}$ where $B_{ijk}$ are ${\cal O}(1)$ coefficients. For $B = 1/3$ however, the trilinear terms do not break $U(1)_R$ symmetry, so $b_{ijk}$ should instead be of order the $U(1)_R$-symmetric soft mass scale.

We first consider the one-loop proton decay channels since, as we will see later, they are less dependent on the UV completion than the tree-level ones. For convenience, we work in the basis where the flavor eigenstates of $ d _{L , i} , d _{ R , i} ^c ,$ and $ u _{R ,i } $ coincide with the mass eigenstates (otherwise we would have additional CKM matrix contributions, which would of course simplify to the same final result), so for instance $y_{d,ij} = \sqrt{2}m_{d,i} \delta_{ij}/v$, where $m_{d,i}$ are the down-type quark masses. We also assume that the quark and squark mass basis are exactly aligned to simplify the index assignments in Fig.~\ref{fig:proton-decay}. Relaxing this assumption complicates the analysis but is not expected to significantly affect our main results. Antisymmetry of $b_{ijk}$ under exchange of $j$ and $k$ (due to $SU(3)$ contraction) further implies that $k=2$ or 3, while kinematic considerations implies $l=1$ or 2 in the left diagram and $m = 1$ in the right diagram. 
For an electron-sneutrino Higgs, we find two decay channels: $uud \rightarrow u \bar{s} \bar{\nu}$ ($p \rightarrow K^+ \, \bar{\nu}$) is the dominant decay channel, while $uud \rightarrow u \bar{u} e^+$ ($p \rightarrow \pi^0 \, e^+$) is subdominant due to CKM suppression, despite having a slight phase space enhancement. (Note that the current bounds on either decay channels are comparable \cite{Nishino:2009aa,Abe:2014mwa}.) Since the dominant decay channel is to the neutrino rather than the charged lepton, the subsequent analysis remains valid in the case of a muon- or tau-sneutrino Higgs.

We now focus on the dominant one-loop channel. Integrating out the gluinos and squarks gives us the standard dimension-6 proton decay operator $\bar{d^c} \bar{u^c} q_L l_L/\Lambda_p^2$. For simplicity we assume that the gluinos are somewhat heavier than the squarks (as is typical in $ R $-symmetric models due to the supersoft mechanism~\cite{Fox:2002bu}) and that $m_{\tilde{g}} \approx m_{\psi_{\tilde{g}}}$. We find that
\begin{equation}
\frac{1}{\Lambda_p^2} \sim \frac{g_s^2}{16\pi^2} \frac{m_{\tilde{g}} b_{112} m_s/v_H}{M_{\tilde{s}_R}^2 M_{\tilde{g}}^2},
\end{equation}
where $g_s$ is the QCD gauge coupling, $m_s$ the strange quark mass, $M_{\tilde{s}_R}$ the mass of the RH strange squark, and $M_{\tilde{g}}$ Dirac gluino mass. We would like to convert the current lower bound of $\Lambda_p \gtrsim {\cal O}(10^{15}) \, \text{GeV}$ \cite{Abe:2014mwa} to an upper bound on $m_{3/2}$. For $B \ne 1/3$, we find that 
\begin{equation}
m_{3/2} \lesssim \left(\frac{1}{c_{\tilde{g}} B_{112}}\right)^{1/2} \left( \frac{M_{\tilde{s}_R}}{1 \, \text{TeV}} \right) \left( \frac{M_{\tilde{g}}}{1 \, \text{TeV}} \right)  \times 0.6 \, \text{keV}.
\end{equation}
We see that for coefficients of order ${\cal O}( 1)$ and sparticle masses of order ${\cal O}( 1) \, \text{TeV}$, we require a gravitino mass of less than ${\cal O}( 1) \, \text{keV}$. In Fig.~\ref{fig:gravitino}, we compare this to the bounds from neutrino masses for different benchmarks of squark and gluino masses. We see in general that the two bounds still remain compatible. 

For $B=1/3$, we instead have
\begin{equation}
m_{3/2} \lesssim \frac{1}{c_{\tilde{g}}} \left( \frac{1 \, \text{TeV}}{b_{112}} \right) \left( \frac{M_{\tilde{s}_R}}{1 \, \text{TeV}} \right)^2 \left( \frac{M_{\tilde{g}}}{1 \, \text{TeV}} \right)^2  \times 4 \times 10^{-7} \, \text{eV}.
\end{equation}
The bound is much stronger in this case, which is not surprising since $U(1)_R$-breaking now only enters once through the Majorana mass insertion and not the trilinear terms. In fact, this bound clearly conflicts with the bounds from neutrino masses, indicating that $B=1/3$ is incompatible with generic gravity-mediated $U(1)_R$-breaking.

Now we move on to the the tree-level channel. Integrating out the squarks to obtain the dimension-6 proton decay operator, we find that
\begin{equation}
\frac{1}{\Lambda_p^2} \sim \frac{a_{112} m_s/v_H}{M_{\tilde{s}_R}^2},
\end{equation}
which translates to a bound of
\begin{equation}
m_{3/2} \lesssim \frac{1}{A_{112}} \left( \frac{M_{\tilde{s}_R}}{1 \, \text{TeV}} \right)^2 \left( \frac{\Lambda}{10 \, \text{TeV}} \right) \times 3 \times 10^{-8} \, \text{eV}.
\end{equation}
This bound is in conflict with the neutrino mass measurements. This suggests either that the $U(1)_R$-breaking is non-generic, or that we require a non-trivial UV completion such that instead of a suppression by $M_{\text{Pl}}\Lambda$ in the tree-level operator, we have an $M_{\text{Pl}}^2$ suppression. In this case we replace $\Lambda$ in the above bound by $M_{\text{Pl}}$, from which we get
\begin{equation}
m_{3/2} \lesssim \frac{1}{A_{112}} \left( \frac{M_{\tilde{s}_R}}{1 \, \text{TeV}} \right)^2 \times 6 \, \text{MeV}.
\end{equation}
which is now consistent with the neutrino constraints and in fact weaker than that from the previous one-loop channel. 

To summarize, we have obtained upper bounds on the gravitino mass $m_{3/2}$ from tree-level and one-loop proton decay channels, assuming generic gravity-mediated $U(1)_R$-breaking. Bounds from both channels are consistent with the bounds from the neutrino mass spectrum, provided that $B \ne 1/3$ and that the tree-level non-renormalizable operator is entirely Planck-scale suppressed. The latter condition implies the need for non-trivial UV completions such that the lighter mass scales $M$ or $\Lambda$ do not enter in the denominator of the tree-level operator, while the suppression is entirely due to $M_{Pl}$. Finally, we emphasize that our entire discussion hinges on the assumption of generic gravity mediation. If $U(1)_R$-breaking is non-generic, certain ${\cal O} (1)$ coefficients may be suppressed or even forbidden.

\section{Conclusions} \label{sect:conclusions}
Supersymmetric models with the Higgs as a slepton are interesting
alternatives to the MSSM. These models have two distinctive features:
an $ R $-symmetry which must be broken by gravity and a mixing of the
Higgs superpartner lepton with the electroweakinos. These properties
allow us to place general bounds on such models from several different
frontiers. In this work, we have studied a variety of such
constraints, which we summarize below.

Previous work has pointed out constraints from neutral and charged
current universality on the mixing of the electron with the gauginos.
These bounds are stringent for the wino, $ M _{\tilde W}  \gtrsim 3.3
\text{ TeV} $, but  weaker for the bino, $ M _{\tilde B} \gtrsim 500
\text{ GeV} $. We revisited these bounds in our framework and compare
them to complementary bounds from low energy probes, which are much
more stringent for the bino, $ M _{\tilde B} \gtrsim 1.2 \text{ TeV} $
and competitive for the wino mass, $ M _{\tilde W} \gtrsim 2.8 \text{
TeV} $. We then moved to study the probing power  a future $ e ^+ e ^-
$ machine. We find large deviations from SM predictions leading to
spectacular reach for such a collider. In particular, for an
integrated luminosity of $ 300 \text{fb} ^{-1} $ and a center of mass
energy of $ 1 \text{ TeV} $, we estimate the potential to probe winos
with masses up to $ 11.5 \text{ TeV} $ in the $ e ^+ e ^- \rightarrow
h Z $ channel.

Higgs-as-slepton models also offer a novel explanation for the
smallness of neutrino masses, arising from spontaneous breaking of the
$U(1)_R$-symmetry due to gravity. We explore the ability of such
models to reproduce the neutrino mass spectrum and the measured mixing
angles. Typically, we find that the models must be UV-completed at a
low scale of at most ${\cal O}( 10 \, \text{TeV})$ in order to
reproduce the large measured mixing angles. Interestingly, this is in
agreement with the scale required to give a sufficiently large top
mass. For the choices $L=0$ and $-1$ (where $L$ parameterizes the
$R$-charge of the non-Higgs-partner leptons), some neutrino mass terms
are not $R$-breaking and hence small neutrino masses require an
additional lepton number symmetry, assumed to be broken at a scale
$M_f$. We find that, under certain assumptions, constraints on the
mixing angles also force $M_f$ to be close to the Planck scale.

Lastly, $ R $-breaking will also generically lead to tree-level proton
decay rates inconsistent with experiment. This puts a restriction on
the type of models which can UV complete the model. Furthermore, we
study loop contributions to proton decay which will be present
regardless of the UV completion. We find that these restrict the
viable range for the gravitino mass to within the range ${\cal O}( 10
\, \text{eV}) - {\cal O}(1 \text{keV})$, which is consistent with the
predictions from neutrino mass measurements. It may be interesting to
study the implications of such a gravitino mass range on observational
cosmology, but we will defer this to future work.

The possibility that the Higgs is the superpartner of the electron is
an intriguing alternative to standard supersymmetric extensions of the
Standard Model. Future tests at the LHC, lepton colliders, low energy
experiments, and of the neutrino mixing patterns each provide an
avenue to discover this variant of supersymmetry.

\acknowledgments{The work of C.B. is partially supported by the Marie
Curie CIG program, project number PCIG13-GA-2013-618439. The work of
JAD is supported in part by NSERC Grant PGSD3-438393-2013. The work of
 JAD, YG and WN is supported in part
by the U.S. National Science Foundation through grant
PHY-0757868 and by a grant from the BSF. C.B. would like to thank the Weizmann Institute for warm hospitality during a visit when this work was initiated.}

\appendix
\section{Feynman rules}
\label{app:FeynmanRules}
In this appendix, we derive the couplings for Yukawa and gauge interactions in the chargino and neutralino mass basis. The mixing matrices used here are derived prior to introducing any $U(1)_R$-breaking.

\subsection{Mixing matrices}
The chargino and neutralino mass matrices are given by
\begin{equation} 
\mathcal{M}_C \equiv \kbordermatrix{
 &e _R^{c,+} &\tilde{W} ^+  & \psi _{\tilde W} ^+  \\ 
e _L ^-  & {\cal O} ( \epsilon _{ \text{NR}} )  &\frac{ g v }{ \sqrt{2} } &0 \\ 
\psi _{\tilde W} ^-  &{\cal O} ( \epsilon _{ \text{NR}} ) & M _{\tilde W}  &0 \\ 
\tilde{W} ^-  & 0& 0& M _{\tilde W}  \\ 
}, 
\hspace{1cm}
\mathcal{M}_N \equiv \kbordermatrix{
 &\tilde{W} ^0  &\tilde{B}   \\ 
\nu _{e,L}   & \frac{ g v }{ 2}& - \frac{ g 'v }{ 2 }  \\ 
\psi _{\tilde W} ^0  &M_ {\tilde{W}} & {\cal O} ( \epsilon _{ \text{NR}} )   \\ 
\psi _{\tilde B}   & {\cal O} ( \epsilon _{ \text{NR}} )&  M _{\tilde B}  \\ 
}, 
\end{equation} 
where $ {\cal O} ( \epsilon _{ \text{NR}} ) $ denotes any non-renormalizable contributions suppressed by the scale $ \Lambda $. While we usually neglect them in our calculations unless specified, we include them here to distinguish them from terms which are identically zero due to $ U(1) _R $ symmetry.

The chargino mass eigenstates are denoted by
\begin{equation} \begin{aligned}
&\begin{pmatrix} \cn{1}\\(\cp{1})^\dagger \end{pmatrix}\mbox{ or }\begin{pmatrix} e_L^{-\prime}\\(e_R^{c,+\prime})^\dagger \end{pmatrix}: &\mbox{mass} &\sim  {\cal O} ( \epsilon _{ \text{NR}} ) ,\\
&\begin{pmatrix} \cn{2}\\(\cp{2})^\dagger \end{pmatrix}: &\mbox{mass} &\approx M_{\tilde{W}} ,\\
&\begin{pmatrix} \cn{3}\\(\cp{3})^\dagger \end{pmatrix}: &\mbox{mass}&\approx M_{\tilde{W}} ,
\end{aligned} \end{equation}
and the neutralino mass eigenstates by
\begin{equation} \begin{aligned}
&\nl{1} \mbox{ or } \nu_{e,L}':  &\mbox{mass} &=0\\
&\begin{pmatrix} \nl{2}\\(\nr{2})^\dagger \end{pmatrix}:  &\mbox{mass} &\approx M_{\tilde{W}} ,\\
&\begin{pmatrix} \nl{3}\\(\nr{3})^\dagger \end{pmatrix}: &\mbox{mass}&\approx M_{\tilde{B}} ,
\end{aligned} \end{equation}
where we have arranged the Weyl fermions into Dirac pairs wherever appropriate.

We denote the unitary transformations between the interaction and mass basis by the matrices $U_{C,L}$, $U_{C,R}$, $U_{N,L}$ and $U_{N,R}$, defined as
\begin{equation} \begin{aligned} 
\left( \begin{array}{c} 
e _L ^-  \\  
\psi _{\tilde{W}} ^-  \\  
\tilde{W} ^-  
       \end{array} \right)
     &= U _{ C,L} \left( \begin{array}{c} 
\cn{1}  \\  
\cn{2}  \\  
\cn{3}
\end{array} \right) , &
\left( \begin{array}{c} 
e _R ^{ c, + }  \\  
\tilde{W} ^+  \\  
\psi _{ \tilde{W} } ^+  
\end{array} \right) &= U _{ C,R} \left( \begin{array}{c} 
\cp{1} \\  
\cp{2} \\  
\cp{3} 
\end{array} \right), \\
\left( \begin{array}{c} 
\nu _{e,L}   \\  
\psi _{\tilde{W}} ^0  \\  
\psi _{\tilde B} 
       \end{array} \right)
     &= U _{ N,L} \left( \begin{array}{c} 
\nl{1}  \\  
\nl{2}  \\  
\nl{3}
\end{array} \right) , &
\left( \begin{array}{c} 
\tilde{W} ^0  \\  
\tilde{B} 
\end{array} \right) &= U _{ N,R} \left( \begin{array}{c} 
\nr{1} \\  
\nr{2} \\  
\nr{3} 
\end{array} \right) .
\end{aligned} \end{equation}  
Note that $\nr{1}$ does not correspond to any fields present in the model and has been introduced simply for notational convenience.

Assuming $ |M _{\tilde W} - M _{\tilde B} | > m _W $, we find that
\begin{equation} \begin{aligned}
U _{ C, L} &= \left( \begin{array}{ccc} 
{\cal O} (1)  & {\cal O} (\epsilon)  & 0 \\  
{\cal O} (\epsilon)  & {\cal O} (1)  & 0 \\  
0 & 0 & 1  
\end{array} \right)  \xrightarrow{{\cal O} ( \epsilon^2 ), {\cal O} ( \epsilon_{\text{NR}}^0 )  } \left( \begin{array}{ccc} 
1 - \epsilon ^2  & \sqrt{2} \epsilon  & 0 \\  
- \sqrt{2} \epsilon  & 1 - \epsilon ^2  & 0 \\  
0 & 0 & 1  
\end{array} \right) , \\ 
U _{ C, R} &= \left( \begin{array}{ccc} 
{\cal O} (1)  & {\cal O} ( \epsilon _{ \text{NR}} )  & 0 \\  
{\cal O} (\epsilon _{ \text{NR}})  & {\cal O} (1)  & 0 \\  
0 & 0 & 1  
\end{array} \right)  \xrightarrow{{\cal O} ( \epsilon^2 ), {\cal O} ( \epsilon_{\text{NR}}^0 )} \left( \begin{array}{ccc} 
1 &0   & 0 \\  
0  & 1 & 0 \\  
0 & 0 & 1  
\end{array} \right) ,\\ 
U _{ N,L} &= \left( \begin{array}{ccc} 
{\cal O} (1)  & {\cal O} (\epsilon)  & {\cal O} ( \epsilon )  \\  
{\cal O} (\epsilon)  & {\cal O} (1)  & {\cal O} ( \epsilon ^2 )  \\  
{\cal O} ( \epsilon )  & {\cal O} ( \epsilon ^2 )  & {\cal O} (1)   
\end{array} \right)  \xrightarrow{{\cal O} ( \epsilon^2 ), {\cal O} ( \epsilon_{\text{NR}}^0 ) } \left( \begin{array}{ccc} 
1 - \epsilon ^2 \frac{1}{2} \left( 1 +  \alpha ^2 t _w ^2 \right)  & \epsilon  & -  \epsilon \alpha t _w     \\  
-  \epsilon  & 1 - \frac{1}{2} \epsilon ^2  & - \frac{  \epsilon ^2 \alpha ^3 t _w  }{ 1 - \alpha ^2 }\\  
 \epsilon \alpha t _w    &  \frac{ \epsilon ^2 \alpha  t _w }{ 1 - \alpha ^2 }  & 1  - \epsilon ^2 \frac{1}{2} \alpha ^2 t _w ^2 
\end{array} \right) , \\ 
U _{ N, R} &= \left( \begin{array}{ccc} 
0 & {\cal O} (1) & {\cal O} (\epsilon ^{ 2})  \\  
0  & {\cal O} (\epsilon ^2 )  & 1
\end{array} \right)  \xrightarrow{{\cal O} ( \epsilon^2 ), {\cal O} ( \epsilon_{\text{NR}}^0 )  } \left( \begin{array}{ccc} 
0  & 1 & - \frac{  \epsilon ^2 \alpha ^2 t _w }{ 1 - \alpha ^2 }  \\  
0 & \frac{  \epsilon ^2 \alpha ^2 t _w }{ 1 - \alpha ^2 } & 1  
\end{array} \right) ,
\end{aligned} \end{equation}
where $ \epsilon \equiv m _W / M _{ \tilde{W} } = g v / (2 M _{\tilde W}) $, $ \alpha \equiv M _{\tilde W} / M _{\tilde B} $ and $ t _w \equiv \tan\theta_w = g'/g$.

\subsection{Couplings for Yukawa interactions}
The Yukawa interactions between the charginos/neutralinos and the Higgs arise from the K\"{a}hler potential of the Higgs/electron supermultiplet. The chargino couplings are given by
\begin{equation} \begin{aligned} 
{\cal L} & \supset - g \frac{ h  }{ \sqrt{2} }e _L^- \tilde{W} ^+ \\ 
& = - \frac{ g }{ \sqrt{2} } h ( U _{ C,L} ) _{ 1i} ( U _{ C,R} ) _{ 2j} \cn{i} \cn{j} .
\end{aligned} \end{equation} 
To ${\cal O} (  \epsilon ) $ and ignoring ${\cal O} ( \epsilon_{\text{NR}})$, this simplifies to
\begin{equation} 
{\cal L} \supset  - \frac{ g }{ \sqrt{2} } h \left( \cn{1} \cp{2} + \sqrt{2} \epsilon \cn{2} \cp{2} \right) .
\end{equation} 

The neutralino couplings are given by
\begin{equation} \begin{aligned} 
{\cal L}  &\supset - g \frac{ h }{ 2} \nu _{e,L} \tilde{W} ^0 + g ' \frac{ h }{ 2} \nu _{e,L} \tilde{B} \\ 
& = - g \frac{ h }{ 2} ( U _{ N,L} ) _{ 1i}  \left[ ( U _{ N,R} ) _{ 1j} - t _w ( U _{ N,L} ) _{ 2j} \right] \nl{i} \nr{j} .
\end{aligned} \end{equation} 
To ${\cal O} (  \epsilon ) $ and ignoring ${\cal O} ( \epsilon_{\text{NR}})$, this simplifies to
\begin{equation} \begin{aligned}  
{\cal L}  \supset - g \frac{ h }{ 2} \left( \nl{1} + \epsilon \nl{2} - t _w \frac{ M _{\tilde W} }{ M _{\tilde B} } \epsilon \nl{3} \right) \left( \nr{1} - t _w \nr{2} \right) .
\end{aligned} \end{equation} 

\subsection{Couplings for gauge interactions}
We begin with the gauge interactions in the interaction basis:
\begin{equation} \begin{aligned} 
{\cal L} & \supset && g \left( \begin{array}{ccc} (\tilde{W} ^+)^\dagger   & (\tilde{W} ^0)^\dagger  & (\tilde{W}^-)^\dagger  \end{array} \right) \left( \begin{array}{ccc} 
W _\mu ^0  & - W _\mu ^+  & 0 \\  
- W _\mu ^-  & 0 & + W _\mu ^+  \\  
0 & + W _\mu ^-  & - W _\mu ^0   
\end{array} \right) \bar{\sigma} ^\mu \left( \begin{array}{c} 
\tilde{W} ^+  \\  
\tilde{W} ^0  \\  
\tilde{W} ^-  
\end{array} \right) \\ 
& &&+ g \left( \begin{array}{ccc}(\psi _{\tilde W}  ^+)^\dagger   & (\psi _{ \tilde{W}} ^ 0)^\dagger  & (\psi _{ \tilde{W}} ^ -) ^\dagger  \end{array} \right) \left( \begin{array}{ccc} 
W _\mu ^0  & - W _\mu ^+  & 0 \\  
- W _\mu ^-  & 0 & + W _\mu ^+  \\  
0 & + W _\mu ^-  & - W _\mu ^0   
\end{array} \right) \bar{\sigma} ^\mu \left( \begin{array}{c} 
\psi _{ \tilde{W}} ^+  \\  
\psi _{ \tilde{W}} ^0  \\  
\psi _{ \tilde{W}} ^-  
\end{array} \right)   \\
& && + g \left( \begin{array}{cc}(\nu _{ e, L}) ^\dagger  & (e _L ^-)^\dagger \end{array} \right) \left( \begin{array}{cc} 
\frac{ W _\mu ^0 }{ 2}  & \frac{ W _\mu ^+  }{ \sqrt{2} }\\  
\frac{ W ^- _\mu }{ \sqrt{2} } & - \frac{ W _\mu ^0 }{2}
\end{array} \right) \bar{\sigma} ^\mu \left( \begin{array}{c} 
\nu _{ e,L}  \\  
e _L ^-  
\end{array} \right)  \\ & &&
 - \frac{g ' }{2} \left( \begin{array}{cc}(\nu _{ e,L}) ^\dagger  & (e _L ^ - )^\dagger \end{array} \right) B _\mu \bar{\sigma} ^\mu \left( \begin{array}{c} 
\nu _{ e,L}  \\  
e _L ^-
\end{array} \right) \\ & &&
+g ' ( e _R ^{ c, +  } ) ^\dagger B _\mu \bar{\sigma} ^\mu e _R ^{ c, + }.
\end{aligned} \end{equation} 
For clarity, we separate this into a few parts before converting to the mass basis.

\subsubsection{Charged current interactions}
The couplings to $ W _\mu ^+ $ are given by 
\begin{equation} \begin{aligned} 
{\cal L} &\supset &&
 g W _\mu ^+ \Bigg\{ ( U _{ N,R} ) _{ 1,i} ^\ast ( \nr{i} ) ^\dagger  \bar{\sigma} ^\mu \cn{3} - ( U _{ C,R} ) _{ 2,i} ^\ast ( U _{ N,R} ) _{ 1j} ( \cp{i} ) ^\dagger  \bar{\sigma} ^\mu \nr{j}  \\
& &&+ \left[ ( U _{ N,L} ) ^\ast _{ 2i} ( U _{ C,L } ) _{ 2j} + \frac{1}{\sqrt{2}} ( U _{ N,L} ) ^\ast _{ 1i} ( U _{ C,L} ) _{ 1j} \right] ( \nl{i} )  ^\dagger \bar{\sigma} ^\mu \cn{j}   \\ 
& && - ( U _{ C,R} ) _{ 3i} ^\ast ( U _{ N,L} ) _{ 2j} ( \cp{i} ) ^\dagger   \bar{\sigma} ^\mu \nl{j} \Bigg\}.
\end{aligned} \end{equation} 
We have used the fact that $\tilde{W}^-$ doesn't mix with $e_L^-$ nor $\psi_{\tilde{W}}^-$ (due to $U(1)_R$ symmetry) to eliminate one of the mixing matrices in the first term. To ${\cal O} (  \epsilon ) $ and ignoring ${\cal O} ( \epsilon_{\text{NR}})$, this simplifies to
\begin{equation} \begin{aligned} 
{\cal L}  & \supset && g W _\mu ^+ \Big[( \nr{2} ) ^\dagger  \bar{\sigma} ^\mu \cn{3} - ( \cp{2}) ^\dagger \bar{\sigma} ^\mu \nr{2} + \frac{1}{\sqrt{2}} (  \nl{1} ) ^\dagger \bar{\sigma} ^\mu \cn{1}    \\ 
& &&  - \frac{1}{\sqrt{2}} \epsilon ( \nl{2} ) ^\dagger \bar{\sigma} ^\mu \cn{1} + ( \nl{2} )  ^\dagger \bar{\sigma} ^\mu \cn{2} +\epsilon  (\cp{3}) ^\dagger \bar{\sigma} ^\mu \nl{1} -  (\nr{3}) ^\dagger \bar{\sigma} ^\mu \nl{2} \Big].
\end{aligned} \end{equation} 
Note that the $V - A$ violating term $ ( \cp{1} ) ^\dagger \bar{\sigma} ^\mu \nl{1} $ does not appear, even when we include higher powers of $\epsilon$ as well as ${\cal O} ( \epsilon _{ \text{NR}} )$. This is not surprising since such a term violates $U(1)_R$ symmetry.

\subsubsection{Neutral current interactions}
We first consider neutral current interactions with the neutralinos, given by
\begin{equation} \begin{aligned} 
{\cal L} & \supset \frac{ g }{ c _w} Z _\mu \frac{1}{2} (\nu _{ e,L}) ^\dagger \bar{\sigma} ^\mu \nu _{ e,L} \\ 
& = \frac{ g }{ c _w } Z _\mu \frac{1}{2} ( U _{ N,L} ) _{ 1i} ^\ast ( U _{ N,L} ) _{ 1j} ( \nl{i} )  ^\dagger \bar{\sigma} ^\mu \nl{j} .
\end{aligned} \end{equation} 
There are no couplings to the photon as expected. To ${\cal O} (  \epsilon ) $ and ignoring ${\cal O} ( \epsilon_{\text{NR}})$, this simplifies to
\begin{equation} \begin{aligned} 
{\cal L} \supset  \frac{ g }{ c _w} Z _\mu \frac{1}{2} & \left\{ ( \nl{1} )  ^\dagger \bar{\sigma} ^\mu \nl{1} + \left[ \epsilon ( \nl{1} ) ^\dagger \bar{\sigma} ^\mu \nl{2}   - t _w \frac{ M _{\tilde W} }{ M _{\tilde B} } \epsilon ( \nl{1} )  ^\dagger \bar{\sigma} ^\mu \nl{3} + \text{h.c.} \right]  \right\}.
\end{aligned} \end{equation} 

Now we move on to the charginos. The couplings to the photon are given by
\begin{equation} \begin{aligned} 
{\cal L} & \supset &&
  e A _\mu \Big[(e _R ^{ c,+} ) ^\dagger \bar{\sigma} ^\mu e _R ^{  c,+} +  (\tilde{W} ^+)^\dagger \bar{\sigma} ^\mu \tilde{W} ^+  + (\psi _{ \tilde{W} } ^+)^\dagger \bar{\sigma} ^\mu \psi _{\tilde W} ^+ \\
 & && - (e _L ^-)^\dagger \bar{\sigma} ^\mu e_L^- - (\psi _{ \tilde{W} } ^-)^\dagger \bar{\sigma} ^\mu \psi _{ \tilde{W} } ^- - (\tilde{W} ^-) ^\dagger \bar{\sigma} ^\mu  \tilde{W}^- \Big]\\
&= &&e A _\mu \left[ ( \cp{i} ) ^\dagger \bar{\sigma} ^\mu \cp{i} - (\cn{i})^\dagger \bar{\sigma} ^\mu \cn{i} \right] .
\end{aligned} \end{equation} 
The couplings are universal as expected since $U(1)_{\text{EM}}$ is unbroken.

The couplings to $Z_\mu$ are given by
\begin{equation} \begin{aligned} 
{\cal L} &\supset 
&& \frac{ g }{ c _w } Z _\mu \left[ (\tilde{W} ^+)^\dagger \bar{\sigma} ^\mu  \tilde{W} ^{ + } - (\tilde{W} ^-)^\dagger \bar{\sigma} ^\mu \tilde{W} ^- + (\psi _{\tilde W} ^+)^\dagger \bar{\sigma} ^\mu \psi _{\tilde W} ^+ - (\psi _{\tilde W} ^-)^\dagger \bar{\sigma} ^\mu \psi _{\tilde W} ^-  - \frac{1}{2} (e _L^-)^\dagger \bar{\sigma} ^\mu e _L ^- \right]  \\ 
& && - \frac{ g }{ c _w } s _w ^2 Z _\mu \Big[  (e _R ^{ c,+} ) ^\dagger \bar{\sigma} ^\mu e _R ^{  c,+} +  (\tilde{W} ^+)^\dagger \bar{\sigma} ^\mu \tilde{W} ^+  + (\psi _{ \tilde{W} } ^+)^\dagger \bar{\sigma} ^\mu  \psi _{\tilde W} ^+ \\
& && - (e _L ^-)^\dagger \bar{\sigma} ^\mu e_L^- - (\psi _{ \tilde{W} } ^-)^\dagger \bar{\sigma} ^\mu \psi _{ \tilde{W} } ^- - (\tilde{W} ^-) ^\dagger \bar{\sigma} ^\mu  \tilde{W}^- \Big]\\
&= 
&& \frac{ g }{ c _w } Z _\mu \Bigg\{ \left[ ( U _{ C,R} ) _{ 2i} ^\ast ( U _{ C,R} ) _{ 2j} + ( U _{ C,R} ) _{ 3i} ^\ast ( U _{ C,R} ) _{ 3j} \right] ( \cp{i}) ^\dagger \bar{\sigma} ^\mu  \cp{j}  \\ 
& && - \left[ \frac{1}{2} ( U _{ C,L}) _{ 1i} ^\ast ( U _{ C,L } ) _{ 1j}+ ( U _{ C,L} ) _{ 2i} ^\ast ( U _{ C,L} ) _{ 2i }+ ( U _{ C,L} ) _{ 3i} ^\ast ( U _{ C,L} ) _{ 3i} \right] ( \cn{i} ) ^\dagger \bar{\sigma} ^\mu  \cn{j}  \Bigg\} \\
& && - \frac{ g }{ c _w } s _w ^2 Z _\mu \left[ ( \cp{i} )  ^\dagger \bar{\sigma} ^\mu  \cp{i} - ( \cn{i} )  ^\dagger \bar{\sigma} ^\mu \cn{i} \right].
\end{aligned} \end{equation}
This comprises of a non-universal part related to mixing between different $SU(2)_L$ representations and a universal part related to $Q$. Using unitarity of $ U _{ C,L} $ and $ U _{ C, R }$, this can be written more succinctly as
\begin{equation} \begin{aligned} 
{\cal L} &\supset && \frac{ g }{ c _w} Z _\mu \Big[ ( 1 - s _w ^2 ) ( \cp{i} )  ^\dagger \bar{\sigma} ^\mu \cp{i} + ( - 1 + s _w ^2 ) ( \cn{i} ) ^\dagger \bar{\sigma} ^\mu \cn{i} \Big]   \\ 
& && + \frac{ g }{ c _w } Z _\mu \Big[ - ( U _{ C,R} ) ^\ast _{ 1i} ( U _{ C,R} ) _{ 1j} ( \cp{i} ) ^\dagger \bar{\sigma} ^\mu \cp{j} + \frac{1}{2} ( U _{ C,L } ) _{ 1i} ^\ast ( U _{ C,L } ) _{ 1j } ( \cn{ i } ) ^\dagger \bar{\sigma} ^\mu \cn{i} \Big].
\end{aligned} \end{equation} 
To ${\cal O} (  \epsilon ) $ and ignoring ${\cal O} ( \epsilon_{\text{NR}})$, this simplifies to
\begin{equation} \begin{aligned} 
{\cal L} & \supset && \frac{ g }{ c _w} Z _\mu \left[ ( 1 - s _w ^2 ) ( \cp{i} )  ^\dagger \bar{\sigma} ^\mu \cp{i} + ( - 1 + s _w ^2 ) ( \cn{i} ) ^\dagger \bar{\sigma} ^\mu \cn{i} \right]   \\ 
& && + \frac{ g }{ c _w } Z _\mu \left\{ -( \cp{1} ) ^\dagger \bar{\sigma} ^\mu \cp{1} + \frac{1}{2} ( \cn{ 1 } ) ^\dagger \bar{\sigma} ^\mu \cn{1} +  \left[ \sqrt{2} \epsilon ( \cn{1} )  ^\dagger \bar{\sigma} ^\mu \cn{2} + \text{h.c.} \right]  \right\}.
\end{aligned} \end{equation} 

\section{Two Higgs Doublet Model}
\label{app:2hdm}
Here we briefly review the Higgs-as-slepton model with two additional Higgs doublets, $ H _u $, $ R _d $. The $ H _u $ can then be used to provide a mass to the top quark, while $ R  _d $ is needed for anomaly cancellation. Table~\ref{tab:model-summary-2} lists the superfields and their gauge and $U(1)_R$ representations.
\begin{table}
\begin{center} 
\begin{tabular}[t]{| c | c | c|}
\hline
 & $ ( SU (3) _C, SU(2) _L ) _Y $ & $U(1)_R$\\
\hline
$H_d \equiv L_e$ & $(1,2)_{-1/2}$ & $0$\\
$E_e^c$ & $(1,1)_1$ & $2$\\
$L_{\mu,\tau}$ & $(1,2)_{-1/2}$ & $1-L$\\
$E_{\mu,\tau}^c$ & $(1,1)_1$ & $1+L$\\
$Q_{1,2,3}$ & $(3,2)_{1/6}$ & $1+B$\\
$U_{1,2,3}^c$ & $(\bar{3},1)_{-2/3}$ & $1-B$\\
$D_{1,2,3}^c$ & $(\bar{3},1)_{1/3}$ & $1-B$\\
$W^{a\alpha}$ & $(8,1)_0 + (1,3)_0 + (1,1)_0$ & $1$\\
$\Phi^{a}$ & $(8,1)_0 + (1,3)_0 + (1,1)_0$ & $0$\\
\hline
$H_u$ & $(1,2)_{1/2}$ & $0$\\
$R_d$ & $(1,2)_{-1/2}$ & $2$\\
\hline
\end{tabular}\end{center} 
\caption{Superfields and their gauge and $U(1)_R$ representations for the 2HDM version of the Higgs-as-sneutrino model.}
\label{tab:model-summary-2}
\end{table}
The most general superpotential consistent with the symmetries (assuming $B \ne 1/3$ and $L \ne 1$) is
\begin{align} 
\mathcal{W} = &\sum_{i,j = 1}^3 y_{d,ij} H_d Q_i D_j^c + \sum_{i,j \in \{\mu, \tau\}} y_{e,ij} H_d L_i E_j^c 
+ \sum_{i,j = 1}^3 y_{u,ij} H_u Q_i U_j^c\nonumber \\
& + \mu H_u R_d + \lambda_S H_u \Phi_{\tilde{B}} R_d + \lambda_T H_u \Phi_{\tilde{W}} R_d \,.
\end{align} 
$ \tilde{h} _u $ and $ \tilde{r} _d $ are now additional neutralinos and charginos which mix with the gaugino and the Higgs-partner lepton. Unlike in the model with the single Higgs doublet, the top quark mass can arise from an $ H _u Q U $ term, removing the need for a low UV cutoff.

For the purpose of deriving the neutrino mass matrix in Sec.~\ref{sect:2hdm-pmns}, after diagonalising the $R$-symmetric terms in the $9 \times 9$ neutralino mass matrix, we now have
\begin{equation}
\nu'_e \simeq \nu_e
+ \left( \frac{M_{\tilde{W}}}{M_{\tilde{B}}} t_w \right) c_\beta \epsilon \, \psi_{\tilde{B}}
- c_\beta \epsilon \, \psi_{\tilde{W}}^0
+ \left( \frac{M_{\tilde{W}}}{\mu} \frac{\lambda_T}{\sqrt{2} g} 
- \frac{M_{\tilde{W}}^2}{M_{\tilde{B}} \mu} \frac{\sqrt{2} \lambda_S t_w}{g} \right) c_\beta s_\beta \epsilon^2 \, \tilde{h}_u^0.
\end{equation}
In contrast to the 1HDM case, $\nu'_e$ now contains a $\tilde{h}_u^0$ component, and some of the coefficients depend on $c_\beta$ and $s_\beta$. The $\tilde{h}_u^0$ component induces the $\nu'_e \nu_\mu$ and $\nu'_e \nu_\tau$ terms in the neutrino mass matrix through the $R$-breaking mass terms $\tilde{h}_u^0 \nu_\mu$ and $\tilde{h}_u^0 \nu_\tau$.

\bibliographystyle{JHEP}
\bibliography{Draft}{}

\end{document}